\documentclass[useAMS,usenatbib,twocolumn]{mn2e}
\usepackage{hyperref}
\usepackage{graphicx}
\usepackage{multirow}
\usepackage{amsmath}
\usepackage{times}
\usepackage{makecell}
\usepackage{caption}

\newcommand{\adsurl}[1]{\href{#1}{ADS}} 
\providecommand{\url}[1]{\href{#1}{#1}}

\title{ A Goldilocks principle for modeling radial velocity noise}
\author[F. Feng et al.]
{F. Feng$^{1}$\thanks{E-mail: f.feng@herts.ac.uk or fengfabo@gmail.com}, M. Tuomi$^{1}$, H. R. A.
  Jones$^{1}$, R. P. Butler$^{2}$ and S. Vogt$^{3}$\\
$^{1}$Centre for Astrophysics Research, University of Hertfordshire, College
Lane, AL10 9AB, Hatfield, UK\\
$^{2}$Department of Terrestrial Magnetism, Carnegie Institute of Washington, Washington, DC 20015, USA\\
$^{3}$UCO/Lick Observatory, Department of Astronomy and Astrophysics, University of California at Santa Cruz, Santa Cruz, CA 95064, USA}
\date{\today}

\begin{document}
\maketitle

\begin{abstract}
The doppler measurements of stars are diluted and
distorted by stellar activity noise. Different choices of noise models and
statistical methods have led to much controversy in the confirmation of
exoplanet candidates obtained through analysing radial velocity
data. To quantify the limitation of various models and methods, we compare
different noise models and signal detection criteria for various
simulated and real data sets in the Bayesian framework. According to our analyses, the
white noise model tend to interpret noise as signal, leading to false positives. On the
other hand, the red noise models are likely to interprete signal as
noise, resulting in false negatives. We find that the Bayesian
information criterion combined with a Bayes factor threshold of
150 can efficiently rule out false positives and confirm true
detections. We further propose a Goldilocks principle aimed at
  modeling radial velocity noise to avoid too many false positives and
too many false negatives. We propose that the noise model with $R_{\rm HK}$-dependent jitter is used in combination with the moving average model to detect planetary signals for M dwarfs. Our work may also shed light on the noise modeling for hotter stars, and provide a valid approach for finding similar principles in other disciplines.
\end{abstract} 
\begin{keywords}
  methods: numerical; methods: statistical; methods: data analysis; Planetary Systems; techniques: radial velocities
\end{keywords}

\section{Introduction}     \label{sec:introduction}
Almost all natural phenomena are studied by collecting and modeling
data, comparing models and inferring model parameters. Since the data
collection and reduction are usually standardised to remove bias and
systematics, the results of data analyses are more influenced by the
choice of models and inference methods than by data
  reduction\footnote{Actually, data reduction is also a kind of
    modeling that converts the primary observations into secondary data
    such as catalogues. But the process is well understood in a theoretical sense. }.

Due to the limited explanation power of theories and models, natural
phenomena are always left with inadequate or incomplete modeling. Thus
our understanding of these unexplained variations (so-called
``noise'') significantly influences how well we can explain data particularly
when noise levels are similar to those of variations. For example, the
glacial-interglacial cycles over the Pleistocene era were probably caused by the orbital variations of the Earth through
modulating the incoming solar radiation \citep{milankovitch30,hays76}. However, this theory is challenged by
various authors \citep{hasselmann76,pelletier97,wunsch04}
using stochastic processes to model the climate change.

In the case of detections of exoplanets in the doppler measurements of
stars, the activity induced radial velocity (RV) variations are called ``excess noise'' or
``jitter'', compared with the RV variations caused by Keplerian
motions of planets (called ``signals''). Jitter is typically
correlated (or red) over various time scales \citep{baluev13}, and
is caused by various mechanisms such as instability of instruments,
magnetic cycles, oscillation, rotation and granulation of stars
\citep{dumusque12}. Jitter actually consists of unmodeled variations
as well as pure noise. This jitter is poorly understood and modeled,
leading to the problem of model incompleteness \citep{fischer16}. To
separate jitter from planetary signals, many noise models are proposed based
either on statistical properties of the RV time series (e.g.,
\citealt{baluev13}) or on astrophysical studies of stellar variability
(e.g. \citealt{rajpaul15}). The number of planetary candidates are
sometimes greatly influenced by the choice of these noise models. For example,
six planets have been claimed to orbit around GJ 581 \citep{vogt09}
based on data analysis applying the white noise model. However, \citet{baluev13}
could not confirm all of them using Gaussian process models. Similar
controversies exist in the confirmation of exoplanets around GJ 667C
using the white noise, moving average and Gaussian process
models \citep{anglada13, feroz13}. These controversies show that the
more flexible the noise model is, the less planetary signals it can
find. We will see this effect in the comparison of noise models. 

Another factor that causes uncertainties in data analysis is the usage
of different statistical methods. For example, many studies claimed
periodicities in the data of mass extinctions and terrestrial impact
craters based on the periodogram or other frequentist approach \citep{alvarez84, raup84, melott12}. But there seems to be no evidence for periodicities in the data based on 
the Bayesian inference \citep{bailer-jones11, feng13, feng14}. The
tension caused by statistics also exists in the confirmation of
planetary candidates even if the same noise model is used. For
example, \cite{tuomi11} only found four planets using Bayesian methods
while \cite{vogt09} found six using the same radial velocity data set
of GJ 581 and the same noise model but based on the periodogram. The solution to such controversies relies on the exploration of appropriate modeling and inference methods based on a better understanding of the mechanisms underlying certain phenomena and a proper choice of statistical tools. 

Most data analysis of radial velocity data was seen based on
frequentist methods, in particular, the Lomb-Scargle periodogram and
adaptations of it, e.g. two dimensional Keplerian Lomb-Scargle periodogram
\citep{otoole09}. However, the periodogram assumes that
the noise in the time series is not correlated and that there is
  only one periodic signal in the data. Despite this, it is misused
to search for multiple periodic signals. For example,  only one Keplerian  component is used to model a superposition of several Keplerian signals. Both of these assumptions are problematic if
the strength of signals is comparable with the noise level \citep{tuomi12c, fischer16}. Furthermore, the periodogram assumes periodicity 
in the data rather than testing it by comparing periodic models with 
other models. Thus periodogram, by definition, is biased in terms of 
model comparison. This is particularly true when the mechanisms 
responsible for certain phenomena are complex and poorly modeled 
(e. g. aperiodic and/or quasi-periodic phenomena). 

Despite these problems, various periodograms are broadly employed to
identify planetary signals in RV observations because they are easy to
calculate. To test the significance of a signal, a selected false alarm probability (FAP) 
of a periodogram is commonly used as a detection threshold. This metric is equivalent 
to the {\it p}-value which is used to reject null hypotheses such as the 
white noise model. However, the choice of null hypothesis is always
arbitrary, and thus makes FAP unable to properly estimate the
significance of a signal. Considering these drawbacks, periodograms should be used 
cautiously particularly in cases when the signal to noise ratio is not 
high and the host star is perturbed by multiple planets
\citep{cumming04,ford07}.

To avoid the above problems of the periodogram, we need a statistical tool to compare models on the same footing rather
than rejecting simple null hypothesis. If we know exactly the underlying physics of certain phenomena, there
would be no need to compare different models. But this is often not
the case for natural phenomena. Hence a proper way to account for
model incompleteness, model complexity and uncertainties of models and
data is crucial for robust data analyses. Fortunately, such inference problems can be
properly dealt with in the Bayesian framework (e.g., \citealt{kass95,
  spiegelhalter02, gregory05, vonToussaint11}). For example, Bayesian
inference methods assess the overall plausibility of a model by
calculating its likelihood averaged over its prior distribution. This approach
naturally accounts for the model complexity and thus models can be
compared properly. 

In addition to an appropriate inference method, a modeling principle should be established through quantifying the limitations of
stochastic and deterministic models. We do this for the RV data of M dwarfs by comparing various noise models in the
following steps. First, we generate artificial data sets using noise
models and the Keplerian model. For these data sets, we compare noise
models using various signal detection criteria. Then we select the
best criterion which confirms most true detections and rejects most false positives. We further apply the criterion to compare models for the data sets with injected Keplerian signals. Based on the results, we quantify the limitations of various noise models and devise a framework of noise models to detect planetary signals. 

Our aim is to provide a quantitative comparison between noise models used in the
literature. We quantify the disadvantages and advantages of each noise
model within the Bayesian framework. Various inference criteria are
investigated for representative RV data sets. We also present a
new principle to model stellar jitter and identify planetary signals. 

This paper is structured as follows. We describe the Bayesian inference
method and signal detection criteria in section
\ref{sec:inference}. In section \ref{sec:models}, we introduce the
models of RV variations, and define their prior distributions.
Then we compare various noise models and signal detection criteria for
artificial data sets in section \ref{sec:artificial}. In section
\ref{sec:injection}, we introduce three RV data sets and inject
planetary signals into them for model comparison. Finally, we discuss the results and conclude in section \ref{sec:conclusions}. 
\section{Data analysis and inference method} \label{sec:inference}
\subsection{Model comparison}\label{sec:comparison}
The Bayesian model comparison relies on the Bayes theorem which is 
\begin{equation}
  P(M_i|D)= \frac{P(D|M_i)P(M_i)}{\sum\limits_{j} P(D|M_j)}, 
\label{eqn:theorem}
\end{equation}
where $P(M_i|D)$ is the {\it posterior} of model $M_i$ for data $D$, $P(D|M_i)$ and
$P(M_i)$ are the {\it evidence} (also called the integrated likelihood) and the {\it prior} of model $M_i$, and the denominator is a normalisation factor. 
Then the ratio of the posteriors of two models is
\begin{equation}
  \frac{P(M_i|D)}{P(M_j|D)} =
  \frac{P(D|M_i)}{P(D|M_j)}\frac{P(M_i)}{P(M_j)}. 
\label{eqn:post_odd}
\end{equation} 
If no model is favoured a priori, i.e. $P(M_i)/P(M_j)=1$, the
posterior ratio becomes 
\begin{equation}
  \frac{P(M_i|D)}{P(M_j|D)} =
  \frac{P(D|M_i)}{P(D|M_j)}\equiv\textrm{BF}_{ij}~,  
\label{eqn:BF}
\end{equation}
where  BF$_{ij}$ is the odds of evidences of model $M_i$ and $M_j$,
and is called {\it Bayes factor}. Following \cite{kass95}, we
interpret $\textrm{BF}_{ij}>150$ as a strong evidence for $M_{i}$ and against $M_j$. 

For model $M$ with parameters $\pmb{\theta}$, the evidence is 
\begin{equation}
  P( D|  M)= \int_{\pmb{\theta}} P( D|\pmb{\theta}, M) P(\pmb{\theta}|M)\textrm{d}\pmb{\theta}~,
  \label{eqn:evidence}
\end{equation}
where $\pmb{\theta}$ is the parameter vector of model $M$, $P(\pmb{\theta}|M)$
is the prior distribution of parameters, and
$\mathcal{L}(\pmb{\theta})\equiv P( D|\pmb{\theta},
M)$ is the {\it likelihood}. The evidence is actually the normalisation
factor of the posterior distribution of model parameters,
\begin{equation}
P(\pmb{\theta}|D,M)  =  \frac{P(D|\pmb{\theta},M)
  P(\pmb{\theta}|M)}{P(D|M)}~.
\label{eqn:post_ dist}
\end{equation}
In most cases, the evidence cannot be calculated analytically due to
the complexity of the likelihood. Thus a Monte Carlo approach is
required either to sample the prior density $P(\pmb{\theta}|M)$ (prior sampling) or to
sample the posterior density $P(\pmb{\theta}|D,M)$ (posterior sampling) or
to sample both. The prior sampling is not appropriate for RV models because the posterior
density always contains multiple modes in the period space related to
planets and activity-induced variations. The modes are typically narrow that the prior
samples may not resolve the posterior distribution properly. Considering these difficulties in prior sampling,
we sample the posterior and calculate the Bayes factors using
various estimators which will be introduced in section \ref{sec:criteria}. 

\subsection{Posterior sampling}\label{sec:posterior}

To sample the posterior density, we use an adaptive
Metropolis-Hastings algorithm of Markov Chain Monte Carlo (MCMC)
developed by \citet{haario01}. This algorithm adjusts the step of the
sampler to explore the posterior efficiently. Considering possible
nonlinear correlation between parameters and a non-Gaussian posterior,
we run adaptive Metropolis-Hastings algorithms to obtain
posterior samples of $10^6-10^7$ for inference. We use the
Gelman-Rubin criteria to judge whether a chain approximately converges to a stationary distribution \citep{gelman92}.

Specifically, we conduct the following steps to produce posterior
samples. First, we run four chains in parallel, and drop out one
half of each chain as ``burn-in'' part. Second, we estimate the so-called
``potential  scale reduction factor'' ($\hat{R}$) by calculating the
variance between and within the chains according to the Gelman-Rubin
criteria. If $\hat{R}$ is less than 1.1, we combine these chains to
provide a statistically representative posterior sample for
inference. Third, we repeat the above two steps to generate chains
with different tempering parameters. A chain is tempered if it is
generated with a probability of move that is proportional to a power of
 the posterior ratio of proposed parameters and current
 parameters. Tempering is used to improve the dynamic properties of a
 chain to explore the whole parameter space efficiently. The chain without tempering is
called ``cold chain'' while the tempered chain is called ``hot
chain''. Considering that the optimal acceptance rate of a Metropolis-Hastings
algorithm is around 0.234 under general conditions \citep{roberts97},
we select the hot chains with acceptance rate between 10\%
and 35\%. Then we identify the potential signal based on the
maximum a posteriori estimation, and use the corresponding parameters
as the initial conditions of a cold chain. Finally, the cold chain
provides a sample drawn from the posterior density of the model. For
models without a Keplerian component, we run cold chains directly to obtain their unimodal
posterior densities. Because we aim at comparing noise models rather
than models with multiple Keplerian components, we only obtain samples
for models with at most one planetary component. 

\subsection{Signal detection criteria}\label{sec:criteria}

Given a statistically representative sample drawn from the posterior 
density, we move on to calculate the evidence using various methods. The integral in
Eqn. \eqref{eqn:evidence} can be calculated by the ``importance
sampling'' method \citep{kass95}, which generates samples from a
density. For example, the harmonic mean (HM) estimator of the evidence
is calculated by averaging the likelihood over samples approximately
drawn from the posterior density. However, this estimator cannot
converge efficiently due to the occasional occurrence of samples with very
low likelihoods. To solve the convergence problem of HM, \citet{tuomi12d}
introduce the truncated posterior mixture (TPM) by drawing samples
from different sections of a MCMC chain to avoid the divergence caused
by low likelihood values. This method is easy to implement because it
only uses the output of Metropolis-Hastings algorithms. But this method is biased if
its free parameter $\lambda$ is large \citep{tuomi12d, diaz16}. In
addition to importance sampling methods, we introduce the one-block
Metropolis-Hastings method developed by \citet{chib01}. The Chib's
estimator (CHIB) is based on the calculation of the posterior of a single point
using samples drawn from the posterior density and the proposal
density of a Metropolis-Hastings sampler. 

Although the evidence can be approximately calculated by the above
methods, they have limitations in applications to complex problems due
to unrealistic assumptions or computation inefficiency (see
\citealt{friel12} and \citealt{han11} for a review). Considering
these, we also introduce various information criteria which are easy
to calculate and thus are frequently used by practitioners. We
introduce three of them: the Akaike Information Criterion (AIC;
\cite{akaike74}), the Bayesian Information Criterion (BIC;
\cite{schwarz78}) and the Deviance Information Criterion (DIC;
\citealt{spiegelhalter02}). The AIC and DIC are criteria motivated from information theory while the BIC is
derived in the Bayesian framework\footnote{Although the BIC is
    derived using the Laplace approximation of a Gaussian likelihood
    distribution and the likelihood distribution in our case is always multimodal,
    we use it because the likelihood is always dominated by the
    Keplerian signal if there is, and the local distribution around the maximum is
    always Gaussian.}. Considering that the sample size of
RV data sets may be small, we use a revised version of AIC introduced
by \citet{hurvich89}. We write the three criteria as follows.
\begin{eqnarray}
AIC&\equiv&-2\ln{\mathcal{L}_{\textrm{max}}}+\frac{2k(k+1)}{N-k-1}\\
BIC&\equiv& -2\ln{\mathcal{L}_{\textrm{max}}}+k\ln{N}\\
DIC&\equiv&D(\bar{\theta})+2p_{D}=\bar{D}(\theta) +p_{D}~,
\end{eqnarray}
where $\mathcal{L}_{\textrm{max}}$ is the maximum likelihood, $k$ is the number of free
parameters\footnote{We assume that a free parameter could be any
  variable in a model as in the case of linear models. Although a more
complex definition of the parameter number could be helpful for
nonlinear models, this is equivalent to changing the Bayes factor threshold which
we will do in section \ref{sec:recover}\,.}, $N$ is the number of data points,
the deviance $D(\pmb{\theta})=-2\ln\mathcal{L}(\pmb{\theta})$ and
the effective number of parameters $p_{D}=\bar{D}-D(\bar{\pmb{\theta}})$. 
To compare the above information criteria with the Bayes factor estimators, we
transform these information criteria into a Bayes factor like quantities \footnote{To make the notation simple, we still use BF to name this quantity.}. It is straightforward to convert the BIC into a Bayes factor because \citet{kass95} argued that $e^{-\Delta BIC_{10}/2} \to
BF_{10}$ when the sample size is large. Here we define
$\Delta\textrm{BIC}_{10}=\textrm{BIC}_1-\textrm{BIC}_0$. We also define Bayes factor using the relative likelihood derived
from AIC, i.e. $BF_{10}=e^{-\Delta\textrm{AIC}_{10}/2}$, where
$\Delta\textrm{AIC}_{10}=\textrm{AIC}_1-\textrm{AIC}_0$. We then
derive Bayes factor from DIC in the same fashion, since the DIC probably
approaches the AIC when parameters are well constrained
\citep{liddle07}.  Note that the transformations from AIC and DIC to Bayes factor
are without theoretical foundation. Rather, it is used to transform
the threshold of AIC or DIC to the threshold of Bayes factor, making AIC or DIC
approximately suitable for Bayesian inference.

With the above evidence estimators and information criteria, we adopt the following
diagnostics for the presence of a Keplerian signal.  
\begin{itemize}
\item The period $P$ of the signal can be constrained from above and below in the posterior density. In other words, it converges to a stationary distribution. 
\item The amplitude $K$ of the signal is significantly greater than
  zero. Specifically, the posterior of $K=0$, i.e. $P(K=0|D,M)$, is
  less than 1\% \footnote{In reality, we fit a normal distribution
  to the posterior sample, and from the best fitted posterior density
  we determine $P(K=0|D,M)$. }. 
\item The evidence of a model with one Keplerian component should be at least 150
  times higher than the evidence for the model without any Keplerian
  component, i.e. BF$_{10}>150$ \citep{kass95}.
\end{itemize}
The above procedure is also used by \citet{tuomi12b} in combination
with the moving average model which we will introduce in the following
section. 

\section{Modeling radial velocity variations} \label{sec:models}
The measured doppler shifts of a star are generated by
gravitational force from star-planet(s) interactions and
stellar activity. The spectroscopic measurements of these doppler shifts yield RV
data with instrument uncertainties and various activity indexes. To
account for these factors, we model the data by combining Keplerian
components and various noise components. In the following
sections, we introduce the basic model which includes the white
noise model and the Keplerian component. Then we add various
noise components onto the basic model to build other models in such a
way that the basic model is nested in the full model given all noise components. 

\subsection{White noise model}\label{sec:basic}
There is good evidence in the architectures of the Solar System and
exoplanetary systems for orbital resonances playing some role (e.g.,
semi-major axes of resonant trans-Neptunian objects). However, the
importance is limited over the typically time span of RV data (e.g.,
\citealt{batygin15}), and so we make the simplifying assumption that
planetary orbits are indenpendent of each other in a planetary
system. Although we only consider at most one Keplerian signal in this
work, we introduce a model of multiple Keplerian signals for general cases. We adopt the following basic model of RV variations,
\begin{eqnarray}
  \hat{v}_b(t_i, \boldsymbol{\theta}) &=& \sum_{j=1}^{n}
                      f_j(t_i)+a\,t_i+b+\sum_{k}c_kI_k\nonumber\\
  f_j(t_i)&=& K_j [\cos(\omega_j + \nu_j(t_i))+e_j\cos(\omega_j)]~,
\label{eqn:basic}
\end{eqnarray}
where $K_j$, $\omega$, $\nu_j$, $e_j$  are the amplitude, the longitude of
periastron, the true anomaly and the eccentricity for the $
j^{\text{th}}$ planetary signal. The true anomaly $\nu$ is an
implicit function of time, and depends on the orbital period $P$ and
the orbital phase at the reference time $M_0$. It can be calculated by
solving the Kepler's equation. Thus the Keplerian component for each
planet contains five free parameters: $K$, $P$, $e$, $\omega$ and
$M_0$. In addition to the above parameters, we use two parameters, $a$
and $b$, to model the acceleration caused either by a companion or by the long period
activity cycles of the star and the reference velocity. We also use
$c_k$ to model the linear dependence of radial velocity on the activity index $I_k$. Specifically, we use $I_F$, $I_B$ and $I_R$ to denote the width
of the spectral lines (FWHM), the bisector span (BIS) and the $log(R'hk)$ ($R_{\rm HK}$),
respectively. Note that these indexes are included into the model
  in a deterministic way. But they will be used in section
  \ref{sec:RJ} and \ref{sec:LRJ} to model the jitter in a stochastic way .

The white noise model accounts for the excess noise through the likelihood function: 
\begin{equation}
\mathcal{L}(\boldsymbol{\theta})\equiv P(D|\boldsymbol{\theta},M_w)=\prod_i
\frac{1}{\sqrt{2\pi(\sigma_i^2+s_w^2)}}\exp\left[-\frac{(\hat{v}_b (t_i,{\boldsymbol\theta}  )-v_i)^2}{2(\sigma_i^2+s_w^2)}\right]
 , 
\label{eqn:like}
\end{equation}
where $\sigma_i$ is the observational noise at time $t_i$, $s_w$ is
the constant amplitude of the white noise, and $v_i$ is the
observed RV at time $t_i$. The jitter depends on stellar activity levels which are partly measured by various shape indexes of spectrum such as
FWHM and BIS and activity proxies such as the $R_{\rm HK}$ index. The noises caused by
activity and instruments are typically correlated \citep{baluev13}, and are too complex
to be modeled deterministically. Thus a range of red noise
models are proposed to remove correlated noises which may mimic
Keplerian signals. In the following sections, we introduce two of
them: the moving average and Gaussian process models. 

\subsection{Moving average}\label{sec:ma}
The moving average (MA) model is used to model the dependence of
current noise on previous noise. The moving average of order $q$ or MA$(q)$ is 
\begin{equation}
  \hat{v}(t_i)=\hat{v}_b(t_i)+\epsilon_{t_i}+\sum_{j=1}^{q}k(t_i,t_{i-j})\epsilon_{t_{i-j}},
\end{equation}
where $k(t_i,t_{i-j})$ is the kernel used to weight the white noise at time
$t_{i-j}$. We introduce two kernel functions, the Laplacian kernel
\begin{equation}
k_L(t_i,t_{i-j}) = w_j \exp(-\beta|t_i-t_{i-j}|) 
\label{eqn:laplacian}
\end{equation}
  and the squared exponential kernel
\begin{equation}
  k_{\textrm{se}}(t_i,t_{i-j}) = w_j\exp(-\beta (t_i-t_{i-j})^2/2) ~,
  \label{eqn:exp_MA}
\end{equation}
where $w_j$ is a positive number for white noise at $t_{i-j}$, and
$\beta$ is a free parameter.

According to \citet{tuomi12}, MA(1) is the best moving
average model which enable the detection of weak signals. In addition,
this model seems to outperform other red noise models in recent RV
Challenge \citep{dumusque16} conducted by Xavier Dumusque \footnote{The details of RV Challenge data sets and results can be found online at \url{https://rv-challenge.wikispaces.com}.}. Hereafter we will use
moving average to denote MA(1) and use the Laplacian kernel if not mentioned
otherwise. 

\subsection{Gaussian process }\label{sec:gp}
The Gaussian process (GP) is included in the RV model by adding
non-diagonal parts to the covariance matrix of the likelihood function (see Eqn. \ref{eqn:like}) which is
\begin{equation}
\mathcal{L}(\boldsymbol{\theta})\equiv
P(D|\boldsymbol{\theta},M_{\text gp})=\frac{1}{\sqrt{|2\pi\mathcal{C}|}}\exp\left[-\frac{1}{2}(\boldsymbol{v}-\boldsymbol{\hat{v}_b})\mathcal{C}(\boldsymbol{v}-\boldsymbol{\hat{v}_b})\right]~,
\label{eqn:gp}
\end{equation}
where $\boldsymbol{v}$ is the observed RV sequence (i.e. $\{v_i\}$), $\boldsymbol{\hat{v}_b}$ is
the RV model expressed by Eqn. \eqref{eqn:basic} (i.e. $\{\hat{v}_b(t_i,\boldsymbol{\theta})\}$), and $\mathcal{C}$ is the
covariance matrix. The covariance matrix is composed of diagonal and
non-diagonal components. The former is related to the Gaussian
measurement noise $\{\sigma_i\}$ and the excess white noise $s_w$, while the latter is determined by a
kernel. To calculate the covariance matrix, we introduce three types
of kernels: Laplacian (L), squared exponential (se) and quasi-periodic
(qp) kernels, which are formulated as follows.
\begin{eqnarray}
  k_{\text L}(t,t')&=&s_r\exp(-|t-t'|/l)~,\nonumber\\
  k_{\text se}(t,t')&=&s_r\exp\left[-\frac{(t-t')^2}{2l^2}\right]~,\nonumber\\
  k_{\text qp}(t,t')&=&s_r\exp\left[-\frac{\sin^2(\pi(t-t')/P_q)}{2l_p^2}-\frac{(t-t')^2}{2l^2}\right]~,
\label{eqn:GP_kernels}
\end{eqnarray}
where $s_r$ is the red noise amplitude, $l$, $l_p$ and $l_e$ are
correlation time scales, and $P_q$ is the period of the quasi-periodic
kernel. The L kernel is used by \citet{baluev13} and \citet{feroz13},
while the se and qp kernels are advocated by \citet{rajpaul15} in
their Gaussian process framework used to disentangle activity-induced variations from planetary signals. In the
comparison of noise models, we will focus on the Laplacian kernel, and
use the other kenels for sensitivity tests. 

\subsection{$R_{\rm HK}$-dependent jitter}\label{sec:RJ}
It is well known that the solar activity is determined by the
magnetohydrodynamic turbulence in the atmosphere, and thus is
difficult to be accurately predicted due to the chaotic nature
of turbulence \citep{petrovay10}. Like the Sun, stars also show
complex activity-induced variations \citep{tobias95} which are partly recorded by
activity indexes in doppler measurements. Therefore the
relationship between RV variations and activity indexes are
probably complex \citep{vanderburg16}, and a deterministic
relationship may not be appropriate to model the activity-induced RV
counterpart. As a result, the dependence of RV on indexes should be modeled statistically. For example, \citet{diaz16} have proposed a linear dependence of
jitter on the $R_{\rm HK}$ index. We call this model ``$R_{\rm HK}$-dependent jitter'' (RJ)
which replace the $s_w$ in Eqn. \eqref{eqn:like} with
\begin{equation}
s(t_i)=s_w+\alpha I_R(t_i)~.
\label{eqn:sj}
\end{equation}
Although \citet{diaz16} used a truncated version of the 
linear function, we don't explore more versions because the $R_{\rm HK}$-dependent jitter model is flexible and representative, and does not require a fine-tuned
threshold to truncate the $R_{\rm HK}$ index. For RV data sets that do not
have $R_{\rm HK}$ index, $I_R$ denotes the activity index which is determined
by measuring the flux at the Ca II H\&K lines with respect to continuum. 

\subsection{Lagged $R_{\rm HK}$-dependent jitter}\label{sec:LRJ}
The activity of a star is determined by the underlying complex and
nonlinear dynamics. Thus the time series generated by stellar activity
typically have long correlation time scale \citep{boffetta99}. In a nonlinear dynamical system, the time series
of state variables are actually projected from the motion on the
manifold of a set of states. According to \citet{takens81} the non-linear state space of the dynamics could
be reconstructed by using only the lagged time series of one variable,
e.g. the $R_{\rm HK}$ index in our case. Thus the jitter in the RV variations can
be modeled as a function of the lagged activity indexes. Considering
that the $R_{\rm HK}$ index is more sensitive to stellar activity \citep{dumusque14}, we let the jitter $s(t_i)$ depend on the previous and subsequent $R_{\rm HK}$ indexes. Then the jitter noise becomes
\begin{eqnarray}
s(t_i) &=& s_w+\alpha I_R(t_i)+f(t_i,t_{i-1}) I_R( t_{i-1})+f(t_i,t_{i+1})  I_R( t_{ i+1})) \nonumber\\
f(t, t') & =&\eta \exp(-\kappa|t-t'|),
\label{eqn:LRJS}
\end{eqnarray}
where $\eta$ is the amplitude of the correlation between jitter, and
$\kappa$ is the inverse of the correlation time scale. Replacing the white noise jitter $s_w$ in Eqn. \eqref{eqn:like} by the above $R_{\rm HK}$-dependent jitter, we define the lagged RJ
model or LRJ. Considering that the lagged $R_{\rm HK}$ may induce the RV counterpart
in a deterministic way, we propose another version of lagged RJ which is
\begin{equation}
  \hat{v}(t_i)=\hat{v}_b(t_i)+f(t_i,t_{i-1}) I_R( t_{i-
    1})+f(t_i,t_{i+1})  I_R( t_{ i+1}). 
\label{eqn:LRJR}
\end{equation}
We call the stochastic version in Eqn. \eqref{eqn:LRJS} LRJ(S) and the
deterministic version in Eqn. \eqref{eqn:LRJR} LRJ(R). They are more
alike than different in terms of accounting for time lagged $R_{\rm HK}$, and
thus share the same name. In other words, the lagged $R_{\rm HK}$ of LRJ(S) is put into the
denominator of the exponential term of the likelihood
(Eqn. \ref{eqn:like}) while the LRJ(R) model contains the lagged $R_{\rm HK}$ in
the numerator. For a simulated data set, only one LRJ version will be chosen
according to the ability of each LRJ model in finding signals. 

In addition to the above mentioned noise models, we combine them to
build compound models to make the comparison more comprehensive. We
combine moving average with $R_{\rm HK}$-dependent jitter to make the MARJ
model, and combine moving average with lagged RJ to make the MALRJ model. For
models with and without one Keplerian component, we add 1 and 0 after the model
name, respectively. For example, MA1 means the moving average model with one
Keplerian component while MA0 means the moving average model without Keplerian
component\footnote{This is not to be mixed with $q^{\mathrm{th}}$ order
 moving average models denoted as MA(q), we only apply MA(1) model. }

\subsection{Prior distributions}\label{sec:prior}
As a necessary part in Bayesian inference, the prior
distributions of parameters are explicitly given for all models in table \ref{tab:prior}. 
For the Keplerian component in the white noise model, we adopt a Jeffreys prior
for the period with a range from one day to the time span of the
RV data sets. Following \citet{tuomi13} we adopt a Gaussian prior
distribution over eccentricity, i.e. $P(e)\propto\mathcal{N}(0,\sigma_e)$ with $\sigma_e=0.2$, to account for observed eccentricity distribution \citep{kipping13} \footnote{Although \cite{kipping13} recommend beta distribution, we adopt the Gaussian
 distribution which is simpler and also flexible enough to describe
 eccentricities.  }. We adopt a uniform prior for the amplitude $K$
with a upper limit of twice the maximum absolute value of RV. This
prior is set not only to speed up the convergence of the chain but
also to account for the fact that the long period signal (longer than
the RV time span) is already modeled as a linear trend in
Eqn. \eqref{eqn:basic}. The jitter amplitude $s$ follows a uniform
distribution from 0 to the upper limit of the prior of $K$. This is also the range of $R_{\rm HK}$-dependent jitter $s( t_i)$. To make the chain
converge quickly, we scale the $R_{\rm HK}$ index such that it has zero mean
and unit variance. For the same reason, we define the ratio of the upper boundary of the prior of $K$, $K_{\text{max}}$, and the difference between the maximum and minimum of the index variation as the
upper limit of the parameters for the dependence of RV on activity indexes, $c_R$, $c_F$ and $c_B$.  For the Gaussian process model, we
find that the likelihood of the Gaussian process model is not sensitive to the time scale $l$, and thus set a narrow boundary for its prior. For the quasi-periodic kernel, we adopt the priors used by \citet{mortier16}.
\begin{table*}
\centering 
\caption{The prior distributions of model parameters. The 
  unit of $c_1$, $c_2$, $c_3$, $\alpha$ and $\eta$ is m/s because the 
  activity indexes are scaled before included in a model. The maximum 
and minimum time of the RV data are denoted by $t_{\textrm{max}}$ and 
$t_{\textrm{min}}$, respectively.}
\label{tab:prior}
  \begin{tabular}  {c*{4}{c}}
\hline 
Parameter&Unit&Prior distribution& Minimum & Maximum\\\hline 
        \multicolumn{5}{c}{\it Each Keplerian signal}\\
$K_j$&m/s&$1/(K_{\text{max}}-K_{\text{min}})$&0&$2|v|_{\text{max}}$\\
$P_j$&day&$P_j^{-1}/\log(P_{\text{max}}/P_{\text{min}})$&1&$t_{\text{max}}-t_{\text{min}}$\\
$e_j$&---&$\mathcal{N}(0,0.2)$&0&1\\
$\omega_j$&rad&$1/(2\pi)$&0&$2\pi$\\
$M_{0j}$&rad&$1/(2\pi)$&0&$2\pi$\\\hline
        \multicolumn{5}{c}{\it Linear trend}\\
$a$&m\,s$^{-1}$yr$^{-1}$&$1/( a_{\text{max}}- a_{\text{min}})$&$-365.24K_{\text{max}}/P_{\text{max}}$&$365.24K_{\text{max}}/P_{\text{max}}$\\
$b$&m/s&$1/( b_{\text{max}}- b_{\text{min}})$&$-K_{\text{max}}$&$K_{\text{max}}$\\\hline
    \multicolumn{5}{c}{\it Index dependence}\\
$c_1$&m/s&$1/( c_{ 1\text{max}}- c_{ 1\text{min}})$&$-c_{ 1\text{max}}$&$K_{\text{max}}/(I_{R\text{max}}-I_{R\text{min}})$\\
$c_2$&m/s&$1/( c_{ 2\text{max}}- c_{ 2\text{min}})$&$-c_{ 2\text{max}}$&$K_{\text{max}}/(I_{F\text{max}}-I_{F\text{min}})$\\
$c_3$&m/s&$1/( c_{ 3\text{max}}- c_{ 3\text{min}})$&$-c_{ 3\text{max}}$&$K_{\text{max}}/(I_{B\text{max}}-I_{B\text{min}})$\\\hline
        \multicolumn{5}{c}{\it Jitter}\\
$s_w$&m/s&$1/(s_{w\text{max}}-s_{w\text{min}})$&0&$K_{\text{max}}$\\
$\alpha$&m/s&$1/(\alpha_{\text{max}}-\alpha_{\text{min}})$&0&$K_{\text{max}}/(I_{R\text{max}}-I_{R\text{min}})$\\
$\eta$&m/s&$1/(\eta_{\text{max}}-\eta_{\text{min}})$&0&$K_{\text{max}}/(I_{R\text{max}}-I_{R\text{min}})$\\
$\kappa$&yr$^{-1}$&$1/(\kappa_{\text{max}}-\kappa_{\text{min}})$&$365.24/(t_{\text{max}}-t_{\text{min}})$&1\\\hline
\multicolumn{5}{c}{\it Moving average}\\ 
$w$&---&$1/(w_{\text{max}}-w_{\text{min}})$&-1&1\\
$\beta$&day$^{-1}$&$1/(\beta_{\text{max}}-\beta_{\text{min}})$&$1/(t_{\text{max}}-t_{\text{min}})$&1\\\hline
\multicolumn{5}{c}{\it Gaussian process}\\
$s_r$&m/s&$1/(s_{r\text{max}}-s_{r\text{min}})$&0&$K_{\text{max}}$\\
$l$&day&$1/(l_{\text{max}}-l_{\text{min}})$&0.01&10\\
$l_p$&day&$l_p^{-1}/\log(l_{p\text{max}}/l_{p\text{min}})$&0.01&100\\
$P_q$&day&$1/(P_{q\text{max}}-P_{q\text{min}})$&1&100\\\hline
  \end{tabular}  
\end{table*}

\section{Model comparison for artificial data sets}\label{sec:artificial}

To ensure that a model is suitable for the data, it is important to
test them using artificial data sets with known noise. Otherwise, the
model or its prior may not be appropriate for certain applications
\citep{fischer16}. We will do this with two types of simulated data
sets, artificial (with artificial signals and noise) and injection (with  artificial signals and real noise) data
sets. Comparing all models for these data sets, we quantify the limitations of
these models and form a signal detection strategy. Fitting different
models to the artificial data sets, we report the signals and models recovered by testing models according to the diagnostics in section \ref{sec:criteria}. 

We generate artificial data sets using three time samples and corresponding measurement errors: the sample from
the Keck measurements of GJ 515A which contains 282 data points and two
subsamples which are generated by randomly drawing 100 time samples from the full
sample of GJ 515A.  We denote these three samples as ``full'', ``sub1'' and ``sub2'', respectively. Then we generate the RV values
using three models: white noise with one Keplerian component (W1),
moving average with one Keplerian component (MA1) and Gaussian process
with one Keplerian component (GP1). We vary the period
$P\in\{20,40,80\}$\,days and the amplitude of constant jitter
$s\in\{0.5,1,2\}$\,m/s. The other parameters are fixed, such that
$K=1$\,m/s, $e=0.1$, $\omega=\pi/2$, $M_0=\pi/2$, $a=-0.1$\,m\,s$^{-1}$yr$^{-1}$,
$b=1$\,m/s, $(w,\beta)=(0.9,0.25~\textrm{day}^{-1})$ and
$(s_r,l)=(1~\textrm{m/s},3~\textrm{day})$. The total number of these
artificial data sets is 54. Then we analyze each data set with the W0,
W1, MA0, MA1, GP0 and GP1 models, and apply the signal detection
criteria (see section \ref{sec:criteria}) to confirm/reject potential signals. The results are shown in table \ref{tab:controled}. 

\begin{table*}
  \begin{center}
\caption{Comparison of noise models for artificial data sets. The
  Bayes factor of the one planet models (e.g. MA1) and corresponding
  zero planet models (e.g. MA0) are calculated for each data
  set. Without applying the threshold of $BF_{10}>150$, the results of
  the test are denoted as A, B and C, which means that the signal is
  correctly recovered, not recovered (false negative) and falsely
  identified (false positive). The results obtained without and with applying the BIC-based BF threshold are written in normal and bold fonts, and corresponding detection numbers (denoted by $N$ and $n$ with subscripts) are reported without and with brackets, respectively. The flag A is underlined if the true model is recovered based on any estimator of the Bayes factor. The numbers of the data sets that have true models selected by any estimator and by the BIC is denoted by $N_T$ and $n_T$, respectively.  } 
\label{tab:controled}
    \vspace{0.05in}
  \begin{tabular}{c| c|*{3}{c}|*{3}{c}|*{3}{c}}
    \hline
\multirow{2}{*}{Sample}  & True model &\multicolumn{3}{ c |}{W1} &\multicolumn{3}{ c|}{MA1}& \multicolumn{3}{ c }{GP1} \\ \cline{2-11}
&Test model&W1&MA1&GP1&W1&MA1&GP1&W1&MA1&GP1\\
    \hline 
    \multirow{6}{*}{full} 
    &$P= $20day; $s=1$m/s&\underline{\bf A}&{\bf A}&{\bf A}&{\bf A}&\underline{\bf A}&{\bf A}&{\bf A}&{\bf A}&B\\
    &$P= $40day; $s=1$m/s&\underline{\bf A}&{\bf A}&{\bf A}&{\bf A}&B&{\bf A}&{\bf A}&{\bf A}&A\\ 
    &$P= $80day; $s=1$m/s&\underline{\bf A}&{\bf A}&{\bf A}&{\bf A}&\underline{\bf A}&A&C&C&B\\ 
    &$P= $20day; $s=2$m/s&\underline{\bf A}&{\bf A}&{\bf A}&{\bf A}&B&B&A&B&B\\ 
    &$P= $40day; $s=2$m/s&B&B&B&A&B&B&{\bf A}&{\bf A}&B\\ 
    &$P= $80day; $s=2$m/s&B&B&B&{\bf A}&\underline{A}&A&B&B&B\\ 
\hline
    \multirow{6}{*}{sub1} 
    &$P=$20day; $s=  0.5$m/s&C&B&B&{\bf A}&\underline{\bf A}&{\bf A}&B&B&B\\ 
    &$P=$40day; $s=  0.5$m/s&\underline{A}&A&A&A&B&B&A&B&B\\ 
    &$P=$80day; $s=  0.5$m/s&\underline{A}&A&A&{\bf A}&B&B&A&A&B\\ 
    &$P=$20day; $s=  1$m/s&B&B&B&B&B&B&{\bf C}&C&C\\ 
    &$P=$40day; $s=  1$m/s&\underline{\bf A}&B&B&B&B&B&A&A&\underline{A}\\ 
    &$P=$80day; $s=  1$m/s&B&B&B&B&B&B&B&B&B\\ 
    \hline
    \multirow{6}{*}{sub2} 
    &$P=$20day; $s=0.5$m/s &\underline{A}&A&A&{\bf A}&\underline{A}&A&B&B&B\\ 
    &$P=$40day; $s=0.5$m/s &\underline{\bf A}&{\bf A}&{\bf A}&{\bf A}&\underline{A}&{\bf A}&{\bf A}&A&\underline{A}\\ 
    &$P=$80day; $s=0.5$m/s &\underline{\bf A}&{\bf A}&A&{\bf A}&\underline{\bf A}&{\bf A}&{\bf C}&C&\underline{A}\\ 
    &$P=$20day; $s=1$m/s &A&B&A&B&B&B&B&B&B\\ 
    &$P=$40day; $s=1$m/s &B&B&B&A&B&{\bf A}&B&B&B\\ 
    &$P=$80day; $s=1$m/s &B&B&B&B&B&B&B&B&B\\\hline  
    \hline
            \multicolumn{11}{c}{\it Numbers of flags for each model
    without (with) applying the threshold of BF $>150$}\\
\hline 
    \multicolumn{2}{c}{$N_A$($n_A$)}&11(7)&9(6)&10(5)& 13(10)&7(4)&9(6)&8(4)&6(3)&4(0)\\
    \multicolumn{2}{c}{$N_B$}&6  &9  &8           & 5   &11   & 9 &7&9&13\\
    \multicolumn{2}{c}{$N_C$($n_C$)}&1(0)  &0  &0           & 0 &0  &0  &3(2)&3(0)&1(0)\\
    \multicolumn{2}{c}{$N_T$($n_T$)}&10(10)&---&---&---&7(3)&---&---&---&3(0) \\
\hline
  \end{tabular}
\end{center}
\end{table*}

\subsection{Comparing models} \label{sec:model_artificial} 

In table \ref{tab:controled}, we observe that the white noise model
recovers signals better than the red noise models. However, the white
noise model also detects more false positives because it interprets
correlated noise as signal. It is a surprise that the moving average model does not
recover itself for the full and sub2 data sets with $(P,K)=$ (40\,day,
1\,m/s). That means the moving average model interprets both the moving average noise and
part of the signal as correlated noise, leading to an underestimation of the significance of the true signal. This indicates that a red noise model may not successfully separate correlated noise from planetary signals.

We also find that the white noise and moving average models could be recovered
through applying at least one Bayes factor estimator for 10 and 7 data
sets, respectively. However, the Gaussian process model is
recovered for only 3 data sets. Due to the flexibility of the Gaussian
process model in fitting a data set, it overfits the noise and thus underfits
the signal, leading to a lower evidence. This problem is also evident from
the fact that the Gaussian process model does not recover any true signals while
the white noise and moving average models recover 4 and 3, respectively, for the Gaussian process generated data sets. 

\begin{figure}
\centering 
  \includegraphics[scale=0.4]{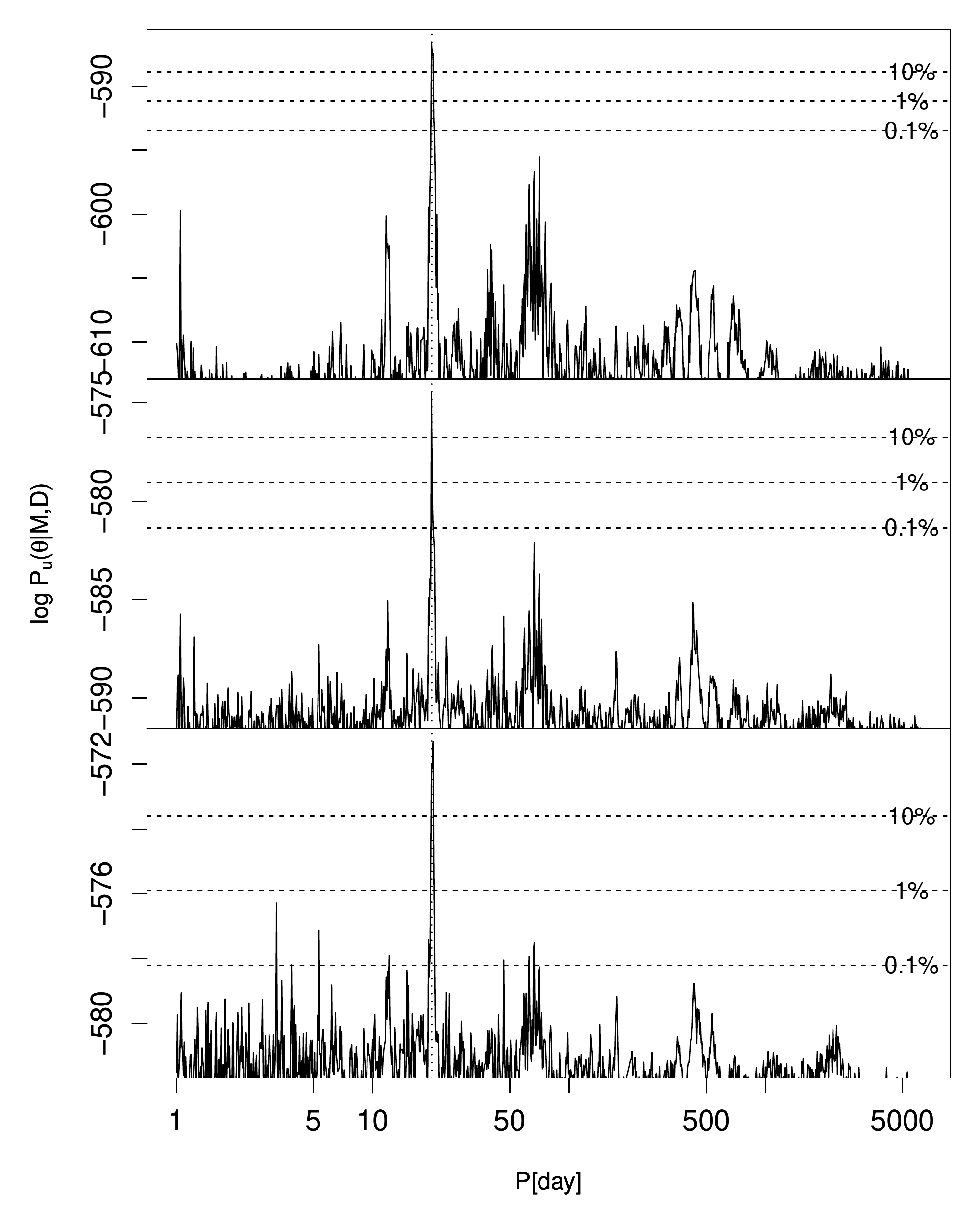}
  \caption{The unnormalised logarithm posterior density from the 
    tempered chains for the white noise (upper), moving average (middle)
    and Gaussian process (lower) models for the full data set
    simulated using the Gaussian process model with $P=20$day and
    $s=1$m/s. To show the significance of signals in each panel, the 10\%, 1\% and 0.1\% of the maximum a posterior is shown by dashed lines. The true period of the artificial signal is denoted by a vertical dotted line.}
\label{fig:post_artificial}
\end{figure}

We can see the above differences between the white noise and moving
average and Gaussian process models from
their posterior densities in figure \ref{fig:post_artificial}. We
calculate the posterior densities by binning the posterior samples
over the period and choosing the maximum posterior in each bin as an
approximation of the marginalised posterior. The significance of the signal
can be inferred from the posterior difference between thresholds and
the difference between the signal and noise. We find that the signal
detected by the white noise model is more significant than those detected by
red noise models, and the signal detected by the Gaussian process model is the
weakest. We also observe that the broad peaks around 70\,days are
probably false positives. The red noise models reduce the significance
of the false positives together with that of the signal. In other words, the
cost of decreasing the false positive rate is increasing the false
negative rate. We also notice that the peak around 1.05\,day is
probably an alias of the signal. But it is not as significant as the
signal for all noise models. In particular, the red noise models seem
to remove the alias efficiently due to their ability of modeling correlated noise over short time scales.  

\begin{figure}
\centering 
  \includegraphics[scale=0.4]{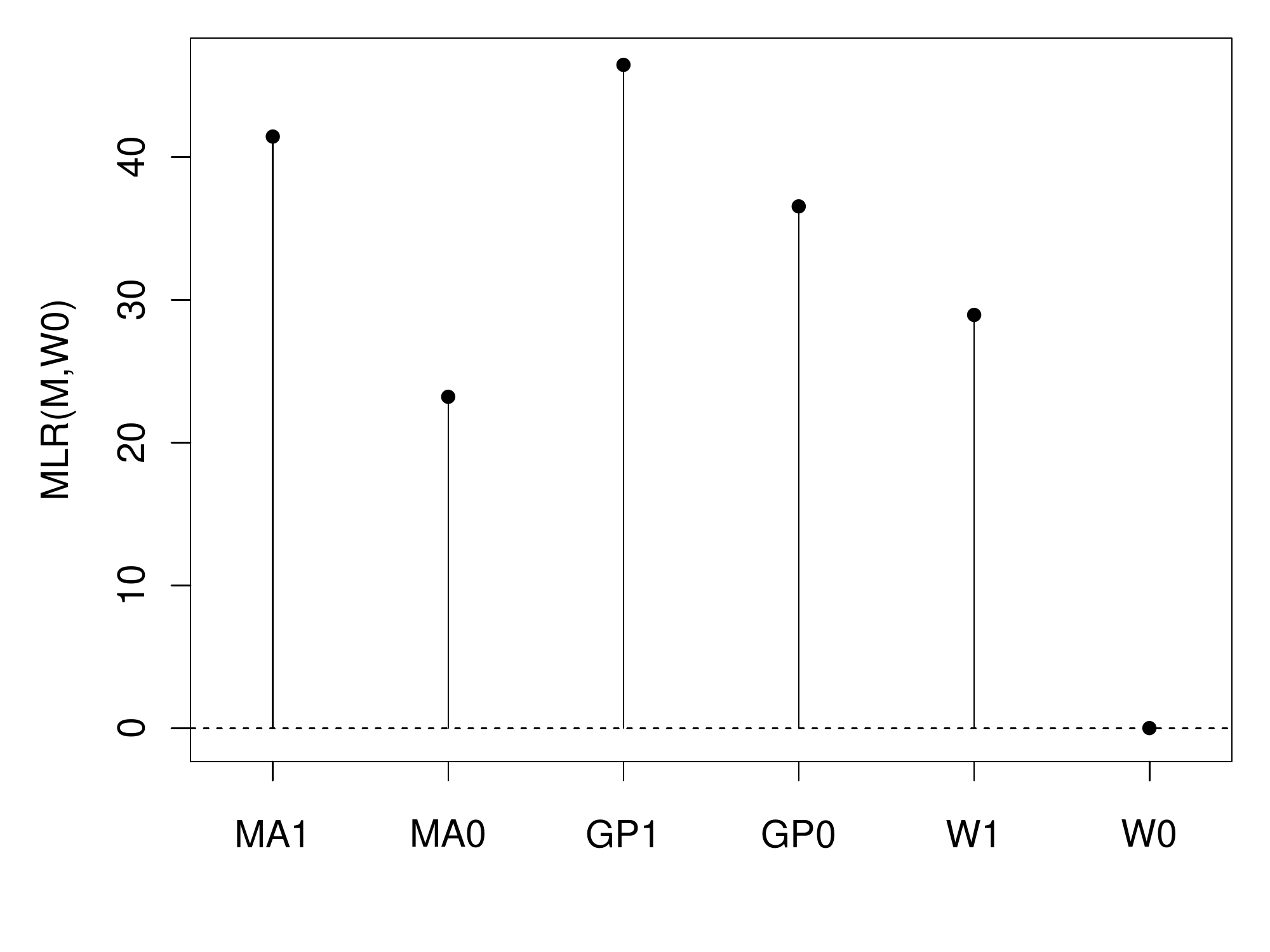}
\caption{The maximum likelihood ratios (MLR) of various models and the
  W0 model for the full data set simulated by the Gaussian process model with
  $P=20$day and $s=1$m/s (the same data used in figure \ref{fig:post_artificial}). }
\label{fig:like_artificial}
\end{figure}

The performance of the models is also revealed by their maximum
likelihoods shown in figure \ref{fig:like_artificial}. We see that the
planetary component can improve the maximum likelihood of white noise model more
significantly than those of other models. This property of the white noise model is generic for all artificial data sets. 

\subsection{Choosing the optimum Bayes factor estimators} \label{sec:estimator_artificial}

\begin{figure}
\centering 
  \includegraphics[scale=0.5]{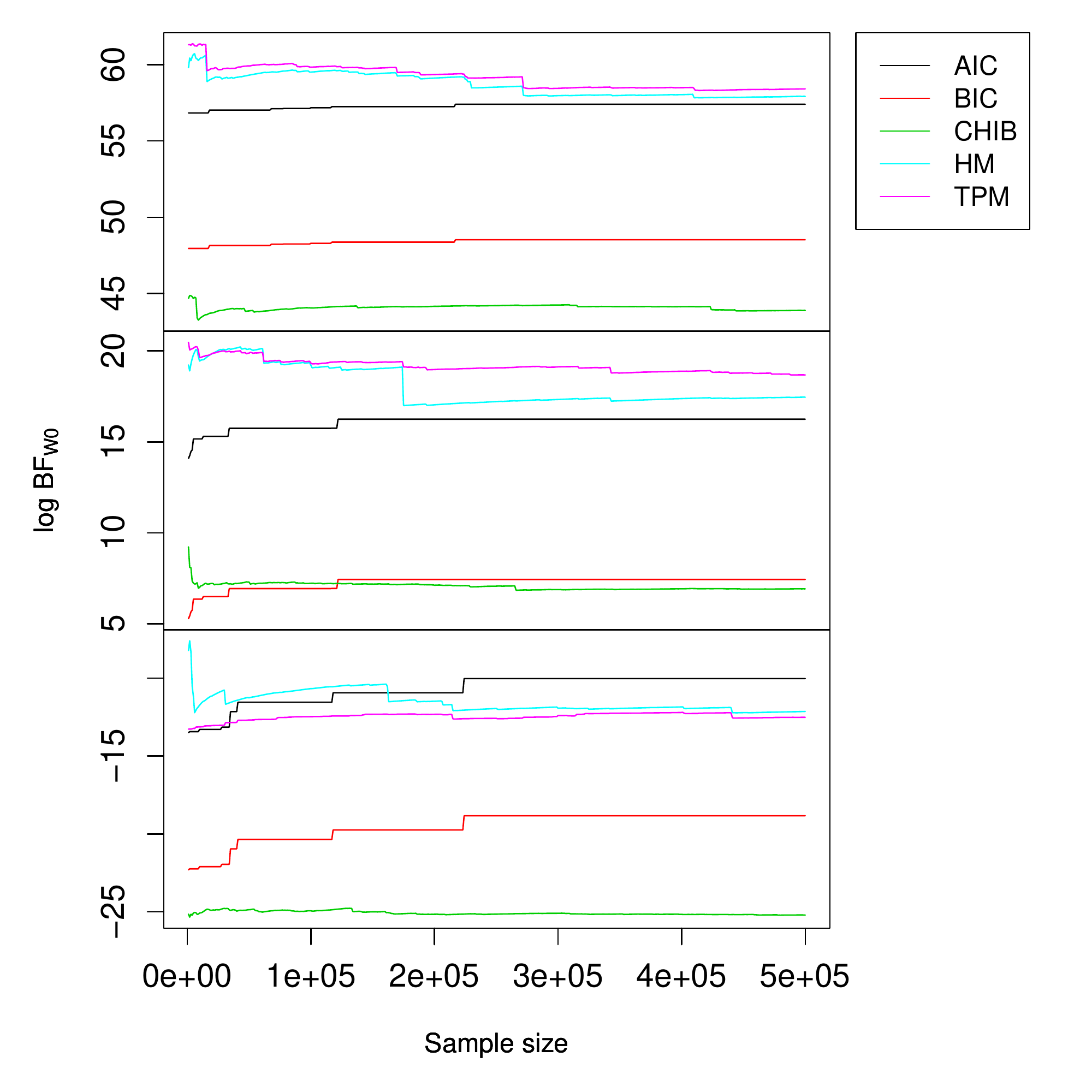}
\caption{The convergence of Bayes factor estimators for the artificial data set
  generated by the moving average model with $P=80$day and $s=1$m/s. The upper, middle and bottom panels show the logarithm Bayes factors of W1, MA1, GP1, with respect to W0, respectively. The TPM estimator of Bayes factor is calculated with $\lambda=10^{-4}$ as recommended by \citet{tuomi12c}. }
\label{fig:BFs_converge}
\end{figure}

\begin{figure}
\centering 
  \includegraphics[scale=0.4]{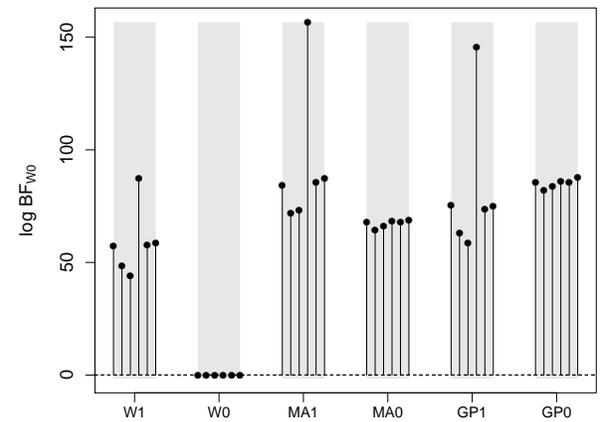} 
  \caption{The Bayes factors estimated by various methods. The heights of the
    lines within each gray bar represent the Bayes factors of a certain model
    with respect to W0 estimated by AIC, BIC, CHIB, DIC, HM and TPM from left to right. }
\label{fig:BFs_stem}
\end{figure}

To further confirm the signal detections in table \ref{tab:controled},
we compare models with and without Keplerian component using the Bayes factor
threshold BF $>150$. We cannot calculate the Bayes factors of models which
do not have any chains converged to the target signal (denoted by flag ``A'' in table
\ref{tab:controled}) due to a lack of statistically representative
posterior samples. For data sets with recovered signals, we calculate
the Bayes factor using the AIC, BIC, CHIB, DIC, HM and TPM estimators. To ensure
the convergence of each method, we increase the size of the posterior
sample gradually and calculate the Bayes factor for each sample size. Since the DIC
cannot converge properly due to the asymmetry and multimodes of the
posterior density, we only show the results for other estimators in figure \ref{fig:BFs_converge}.
We find that the HM and TPM estimators give similar results since the HM is
just a special case of the TPM \citep{tuomi12d}. However, neither of them
converge very well due to the occasional occurrence of samples with
very low likelihoods. The Bayes factor estimated by AIC is always
higher than that estimated by the BIC and CHIB. We also see these
differences from the Bayes factors calculated using the full posterior
sample in figure \ref{fig:BFs_stem}. For models
with one Keplerian component, we find that the DIC always estimate
much higher Bayes factors while the BIC and CHIB estimate the lowest
Bayes factors. However, for models without any Keplerian component, all methods
give simliar Bayes factors, which justifies the usage of each method in the
cases of unimodal posterior densities.  We further investigate all data
sets, and find that the Bayes factors estimated by AIC, HM and TPM are comparable, and the BIC and CHIB estimators always give similar results. 

We then test the significance of signals using the Bayes factor threshold of
150. For each estimator of the Bayes factor, we apply the threshold to confirm
recovered signals (denoted by `` A''), false positives (denoted by ``
C'') and recovered models (denoted by `` T''). The ratios of
confirmed and total number of them for all data sets\footnote{To keep the notation simple, we continue to use $N$ and $n$ with subscripts to denote them (see table \ref{tab:controled}).} are used to characterize the ability of estimators combined with the Bayes factor threshold in confirming true signals and noise properties. The results for all estimators are reported in table \ref{tab:ratio}. 
\begin{table}
  \begin{center}
    \caption{The ratios used to characterise estimators. The notations are similar to those in table \ref{tab:controled}. }
\label{tab:ratio}
  \begin{tabular}{c| *{6}{c}}
&AIC&BIC&CHIB&DIC&HM&TPM\\
    \hline 
$n_A/N_A$  &0.96&0.58&0.47&1.0&0.90&0.90\\
$n_C/N_C$  &0.75&0.25&0.13&1.0&0.88&0.88\\
$n_T/N_T$  &0.75&0.65&0.60&0.40&0.70&0.50\\
  \end{tabular}
\end{center}
\end{table}

In this table, we find that the DIC is not appropriate for confirming 
detections because it cannot rule out any false positive. On the 
contrary, the CHIB estimator is able to get rid of almost all false positives, but 
can only confirm 47\% true detections. An appropriate choice is the AIC 
which could rule out about one quarter of the false positives, and 
confirm 96\% of the true detections. Although the AIC is not a
Bayesian criterion, we regard the AIC combined  with a threshold as a
practical tool to confirm detections. In the case of exoplanet detection, avoiding false 
positives is more important than avoiding false negatives. Such 
requirements can be satisfied by the BIC which rules out 75\% false 
positives and confirms 58\% true detections. In addition, the BIC recovers 75\% true
models while the CHIB method recovers 60\%, justifying our choice of
the BIC rather than the CHIB estimator to rule out
false positives. Although the HM and TPM estimators confirm most true
detections, they have convergence problems as we have mentioned (see
figure \ref{fig:BFs_converge}). 

In summary, the white noise model is able to detect weak signals
efficiently while the Gaussian process are so flexible that signals are
interpreted as correlated noise. The performance of the moving average model is somewhere between the performance of Gaussian process and white noise models. Moreover, most false positives could be ruled out by the BIC-estimated Bayes factor threshold of 150. 

\section{Model comparison for injection data sets}\label{sec:injection}

In the above section, we have analysed the artificial data sets with
known noises and signals. To make our analysis more general, we apply
the same analysis method to data sets with known signals but with
noises from real data. We adopt three data sets, the HARPS measurements of
GJ 1 (44 epochs) and GJ 361 (101 epochs), and the Keck
measurements of GJ 445 (64 epochs). For each data set, we use the RVs,
measurement errors and activity indexes of $R_{\rm HK}$. We also consider the
RV dependence on the FWHM and BIS index (see Eqn. \ref{eqn:basic}) if
they are available. Considering that the data of GJ 445 is not
published before, we show it in the appendix. We introduce the three targets in the following section. 

\subsection{Radial velocity data}\label{sec:data}
Most nearby stars now have precision radial velocities recorded for 
them. As of 2016 April the ESO archive for the HARPS instrument finds 
over 7000 different targets although most only have a few epochs some 
objects have large numbers, e.g., nearly 20,000 for alpha Centauri 
B. We focus our attention on radial velocity data for nearby M 
dwarfs. We choose these targets because they appear to have relatively 
lower activity noise. We have attempted to
focus on targets which have reasonable sampling, good precision and
for which there is enough radial velocity data to make detections but
where there is not a strong known signal. We have drawn these from the
sample of \citet{tuomi16} and specifically choose to study data from both HARPS \citep{mayor03} and HIRES \citep{vogt94}. These are two of the pre-eminent radial velocity instruments whose design, calibration and processing can be considered both reliable and independent. 

GJ 1 is a metal-poor (e.g., [Fe/H]=-0.45; \citealt{neves12}) M2 dwarf
at a distance of 4.5\,pc \citep{leeuwen07} without any reported
planetary companions
(e.g. \citealt{zechmeister09}). \cite{mascareno15} find that it has a
rotation period of $60.1\pm5.7$\,d based on spectral activity
indexes. Analysis of the ASAS photometric data \citep{pojmanski02} does not
confirm this rotation period \citep{tuomi16} though there are relatively modest 42 photometric data points spanning less than a year. We consider the 44 HARPS epochs which we have extracted from the ESO archive. 

GJ 445 is a metal-poor (e.g., [Fe/H]=-0.30; \citealt{neves12}) M4 dwarf
at a distance of 5.4\,pc \citep{leeuwen07} without any reported
planetary companions or rotational signals. Here we consider 64 epochs
of Keck data. 

GJ 361 is a slightly metal-poor (e.g., [Fe/H]=-0.11; \citealt{neves12})
M1.5 dwarf at 11\,pc \citep{leeuwen07}. \cite{tuomi16} find a low
amplitude signal (3.82m/s) with a period of 28.9\,d which is removed
from the data. \cite{tuomi16}  do not find any evidence for significant periodicities in 241 ASAS V-band photometric observations of the star spanning 2,298 days. We utilise 101 epochs of HARPS radial velocity data. 

To extract the noise from the GJ 361 data set, we analyze the data of
GJ 361 with the moving model and identify the signal, and subtract the
signal from the data set. This subtracted version is called ``GJ361
subtracted''. To ensure that no significant signal exist in the GJ 1, GJ 445 and GJ 361 subtracted data sets, we fit all models in section \ref{sec:models} to them and do not find 
any significant signal which satisfies the signal detection 
criteria. Fitting the W0 model to all data sets, we obtain the 
posterior of jitter-induced white noise $s_w$, and use the mean value
as a reference point for choosing the amplitudes of injected Keplerian signals. 

\subsection{Recovering signals}\label{sec:recover}

To inject signals, we vary the amplitude and period of the Keplerian
component and keep other parameters fixed. The amplitude of a signal is
varied in such a way that the signal strength is lower, comparable or
higher than the jitter level. We adopt $P\in\{20,40,80\}$\,day for all data sets,
$K\in\{1,2,4\}$\,m/s for GJ 1 and GJ 361, and $K\in\{4,6,8\}$\,m/s for GJ
445. The other Keplerian parameters are set $e=0.1$, $\omega=\pi/2$,
$M_0=\pi/2$. Finally, we fit all noise models with and without planet
to the injection data sets, and report the detections confirmed by the
signal detection criteria in table \ref{tab:injection}. 

\begin{table*}
  \begin{center}
\caption{Model comparison for RV data sets without and with injected signals. We use the LRJ(S) and MALRJ(S) models to fit the GJ1 data set, and
  apply the LRJ(R) and MALRJ(R) models to fit the other data sets.  The meanings of A,
  B and C are described in table \ref{tab:controled}.}
\label{tab:injection}
  \vspace{0.05in}
  \begin{tabular}{c| c| *{7}{c}}
    \hline
 data set&injected signal&MA1& GP1&W1& RJ1& LRJ1& MARJ1& MALRJ1\\
    \hline
    \multirow{10}{*}{GJ1}
   &---&B&B&B&B&B&B&B\\
   &P=  20 day, K=  1\, m/ s   &B&B&B&B&B&B&B\\
   &P=  20 day, K=  2\, m/ s  &B&B&{\bf A}&{\bf A}&A&A&A\\
   &P=  20 day,  K=  4\, m/ s   &{\bf A}&{\bf A}&{\bf A}&{\bf A}&{\bf A}&{\bf A}&{\bf A}\\
   &P=  40 day,  K=  1\, m/ s   &B&B&C&B&B&B&B\\
   &P=  40 day,  K=  2\, m/ s   &A&B&{\bf A}&{\bf A}&{\bf A}&A&{\bf A}\\
   &P=  40 day,  K=  4\, m/ s   &{\bf A}&{\bf A}&{\bf A}&{\bf A}&{\bf A}&{\bf A}&{\bf A}\\
   &P=  80 day,  K=  1\, m/ s   &B&B&C&B&C&B&B\\
   &P=  80 day,  K=  2\, m/ s   &A&B&{\bf A}&{\bf A}&{\bf A}&{\bf A}&A\\
   &P=  80 day,  K=  4\, m/ s   &{\bf A}&{\bf A}&{\bf A}&{\bf A}&{\bf A}&{\bf A}&{\bf A}\\
\hline
    \multirow{10}{*}{GJ445} 
   &---&B&B&C&B&B&B&B\\
   &P=  20 day,  K=  4\, m/ s   &B&B&B&B&B&B&B\\
   &P=  20 day,  K=  6\, m/ s   &B&B&C&C&A&B&B\\
   &P=  20 day,  K=  8\, m/ s   &{\bf A}&A&{\bf A}&{\bf A}&{\bf A}&{\bf A}&{\bf A}\\
   &P=  40 day,  K=  4\, m/ s   &A&B&{\bf C}&B&{\bf A}&B&A\\
   &P=  40 day,  K=  6\, m/ s   &A&B&{\bf A}&{\bf A}&C&A&A\\
   &P=  40 day,  K=  8\, m/ s   &A&A&{\bf A}&{\bf A}&{\bf A}&{\bf A}&{\bf A}\\ 
   &P=  80 day,  K=  4\, m/ s   &B&B&{\bf C}&C&B&B&B\\
   &P=  80 day,  K=  6\, m/ s   &B&B&B&B&B&B&B\\
   &P=  80 day,  K=  8\, m/ s  &B&B&{\bf A}&{\bf A}&{\bf A}&B&B\\
    \hline
    \multirow{10}{*}{GJ361 subtracted} 
    &---&B&B&C&C&C&B&B\\
    &P=  20 day,  K=  1\, m/ s &B&B&B&B&B&B&B\\
    &P=  20 day,  K=  2\, m/ s &B&B&B&B&B&B&B\\
    &P=  20 day,  K=  4\, m/ s &{\bf A}&{\bf A}&{\bf A}&{\bf A}&{\bf A}&{\bf A}&{\bf A}\\
    &P=  40 day,  K=  1\, m/ s &B&B&C&B&C&B&B\\
    &P=  40 day,  K=  2\, m/ s  &{\bf A}&{\bf A}&{\bf A}&{\bf A}&{\bf A}&A&A\\
    &P=  40 day,  K= 4\, m/ s  &{\bf A}&{\bf A}&{\bf A}&{\bf A}&{\bf A}&{\bf A}&{\bf A}\\
    &P=  80 day,  K=  1\, m/ s &B&B&B&C&B&B&B\\
    &P=  80 day,  K=  2\, m/ s &B&B&{\bf A}&{\bf A}&{\bf A}&B&A\\
    &P=  80 day,  K=  4\, m/ s  &{\bf A}&{\bf A}&{\bf A}&{\bf A}&{\bf A}&{\bf A}&{\bf A}\\
    \hline
            \multicolumn{9}{c}{\it Numbers of flags for each model
    without (with) applying the threshold of BF $>150$}\\
\hline 
         \multicolumn{2}{c}{$N_A$($n_A$)}&13(8)&9(7)  &15(15)&15(15)  &16(14)&13(9)&15(9)\\
     \multicolumn{2}{c}{$N_B$}    &17  &21  &7  &11      &10   &17  &15\\
   \multicolumn{2}{c}{$N_C$($n_C$)}      &0  &0  &8(2) &4(0)  &4(0)  &0  &0\\
\hline
  \end{tabular}
\end{center}
\end{table*}

As seen from table \ref{tab:injection}, no strong signals are found by
  any noise model, although the W1, RJ1 and LRJ1 models seem to
  identify weak signals which fail to pass the Bayes factor threshold. The
table shows that the LRJ1 model detects the most signals without
applying the BIC-estimated Bayes factor threshold (BF$_{10}>150$) while the W1 and RJ1 models find the most signals once applying the
threshold. On the contrary, red noise models only recover less than
half of the injected signals and even less if applying the Bayes factor
threshold, implying that they are much more conservative and prone to
false negatives. However, if without applying the Bayes factor
threshold, the moving average model can recover 13 signals. As a red noise model, moving average is not as
flexible as Gaussian process, and thus is able to identify more signals. In
addition, most true signals recovered by the W1, RJ1 and LRJ1 models
are also strong in the posterior densities of the moving average model, although
they may not satisfy the detection criteria. On the contrary, the
false positives are never strong in the posterior distributions of
moving average. Hence the moving average model can be used to confirm true detections and reject false ones. 

To ensure that the results are not sensitive to the choice
of kernels, we adopt the squared exponential kernel for the moving average models
(see Eqn. \ref{eqn:exp_MA}), the squared exponential and
quasi-periodic kernels for the Gaussian process models (see
Eqn. \ref{eqn:GP_kernels}). We fit the red noise models with these new
kernels to the GJ1 data set with $(P,K)=$ (20\,day, 2\,m/s), the GJ445 data set with
$(P,K)=$ (80\,day, 8\,m/s), and the GJ361 subtracted data set with
$(P,K)=$ (80\,day, 2\,m/s). For all these three data sets, we don't
find any statistically significant improvement by changing kernels. 

\begin{figure}
  \centering
  \includegraphics[scale=0.4]{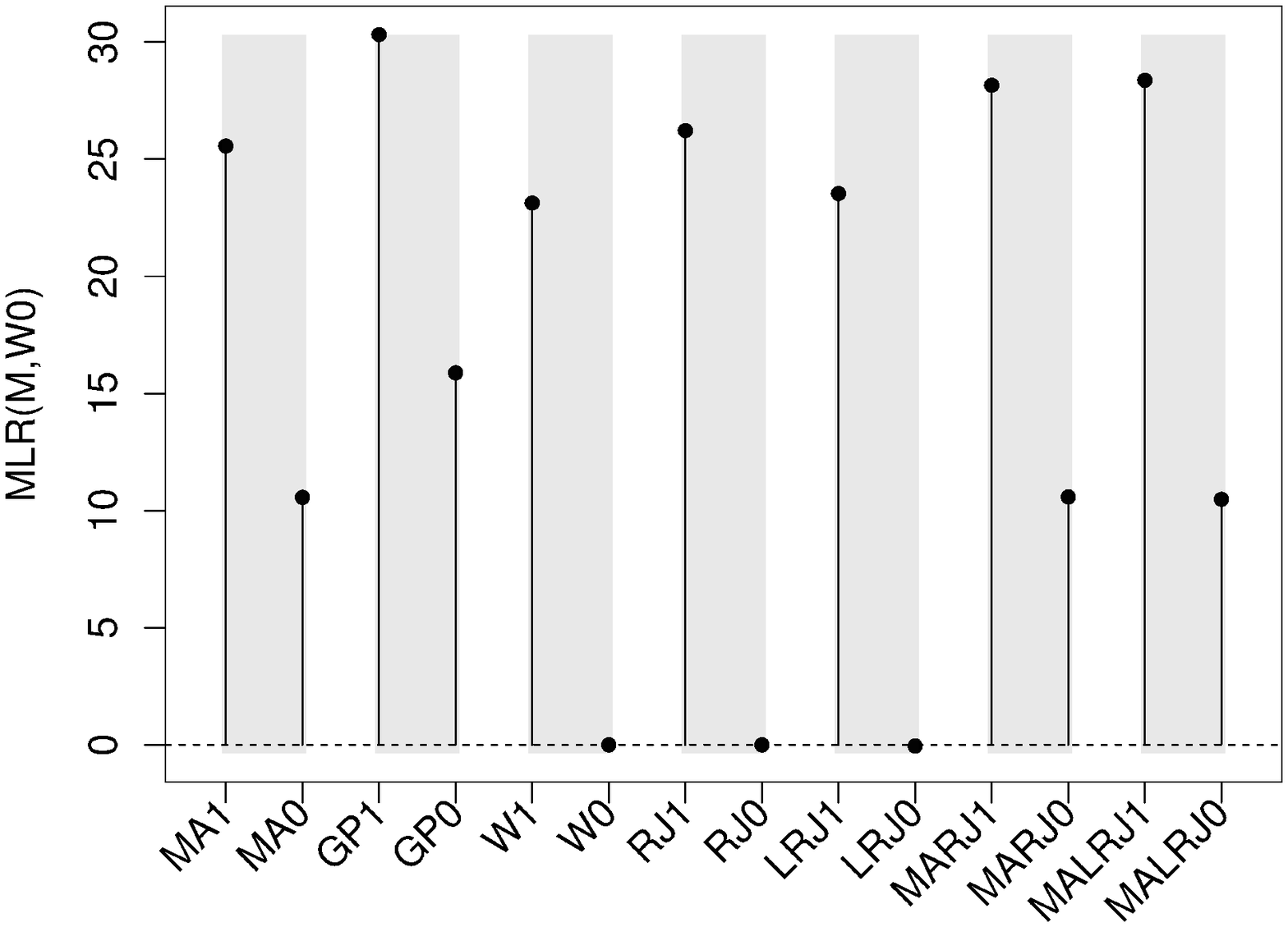}
  \includegraphics[scale=0.4]{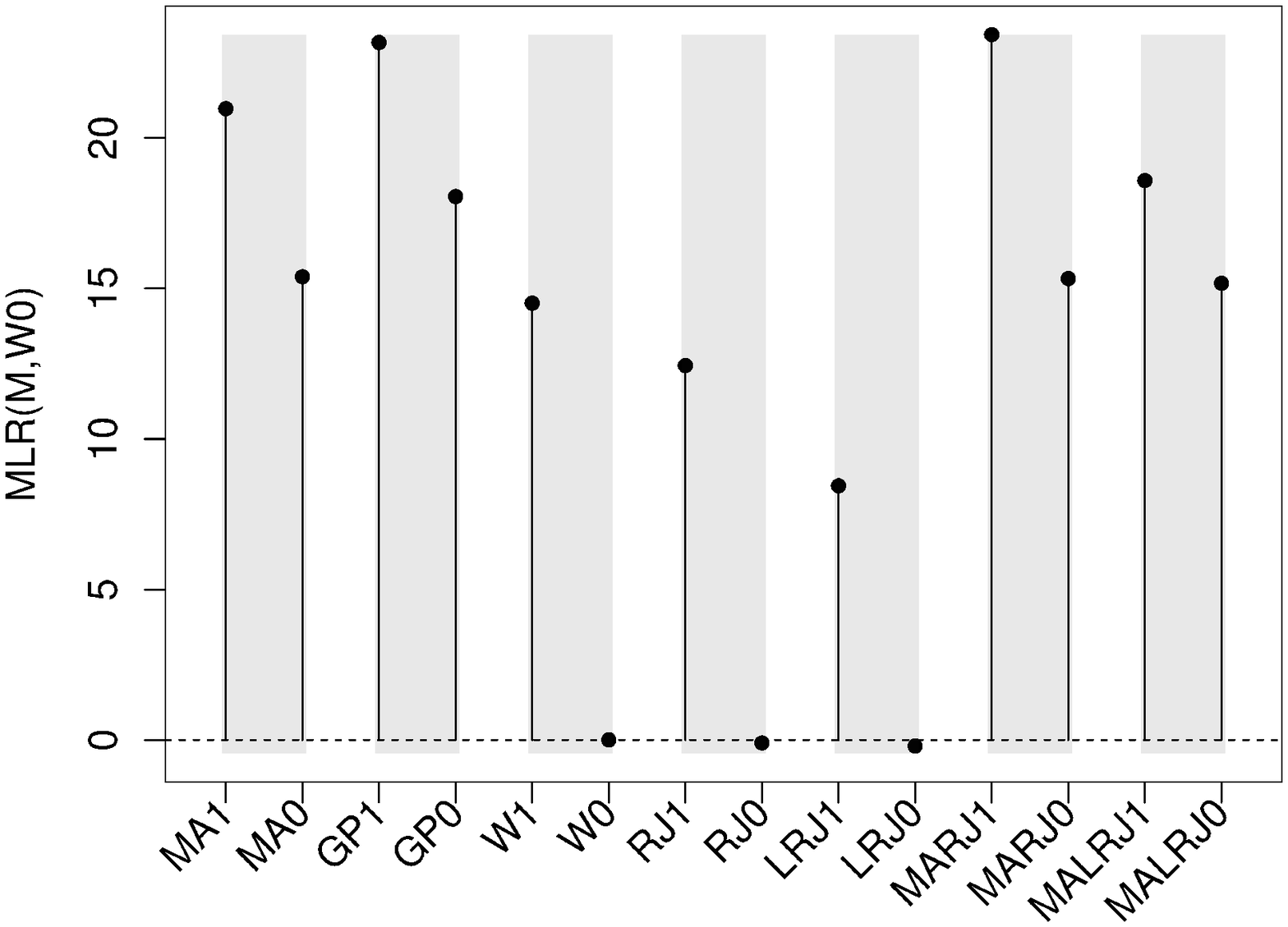}
  \caption{The maximum likelihood ratio (MLR) of noise models and the W0
    model for the GJ445 data set with $P=40$\,day and $K=8$\,m/s and
    the GJ361 subtracted data sets with $P=80$\,day and
    $K=2$\,m/s. Each grey bar encloses the MLRs for one noise model
    with and without planetary component. }
 \label{fig:like_injection}
\end{figure}

As we have mentioned in section \ref{sec:artificial}, the red noise
models interprets signals as noise, and thus adding a Keplerian component to
a red noise model would not improve the likelihood as much as the
W1 model does. This is evident from the comparison of the maximum
likelihoods of all models in figure \ref{fig:like_injection}. We observe
that the likelihoods of W1, RJ1 and LRJ1 are much higher than those of
W0, RJ0 and LRJ0, indicating the necessity of adding one Keplerian component into the
noise model. On the contrary, the Keplerian component does not significantly improve
the likelihoods of the red noise models, in particular the Gaussian process model. 

\begin{figure*}
\centering
  \includegraphics[scale=0.6]{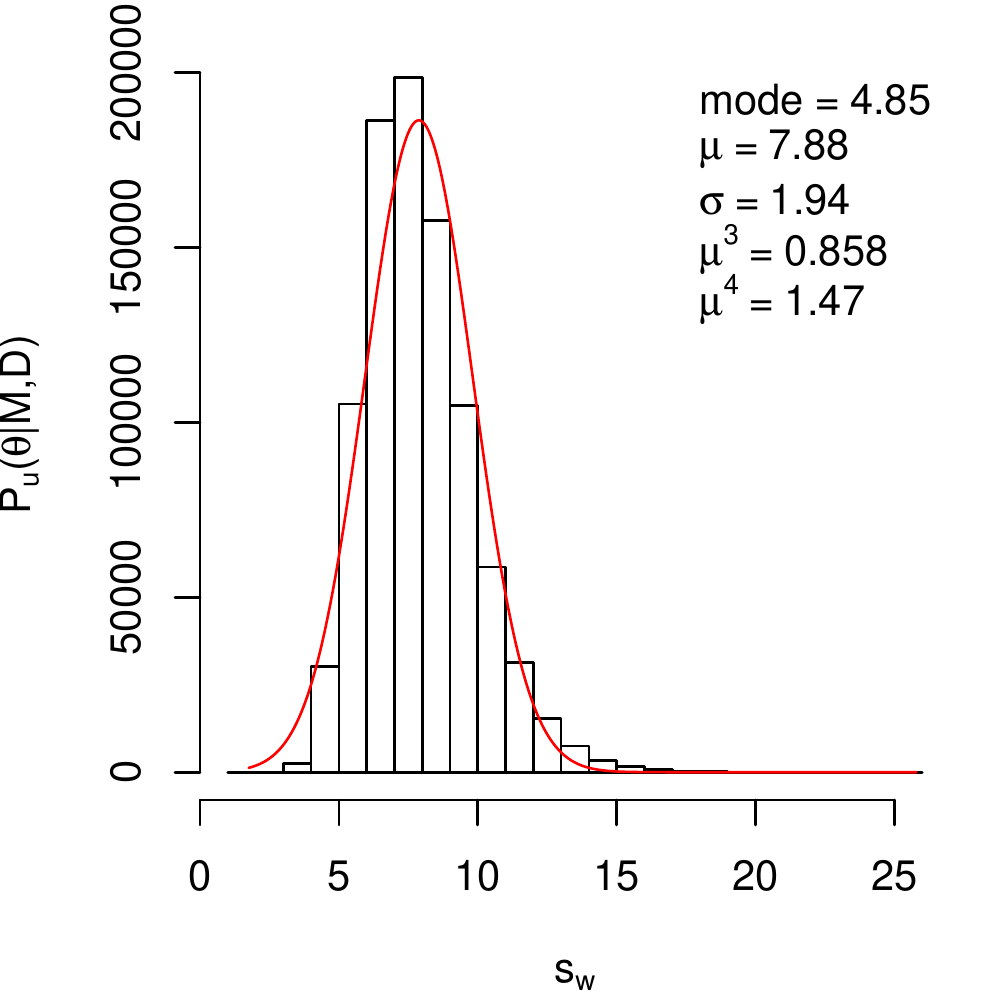}
  \includegraphics[scale=0.6]{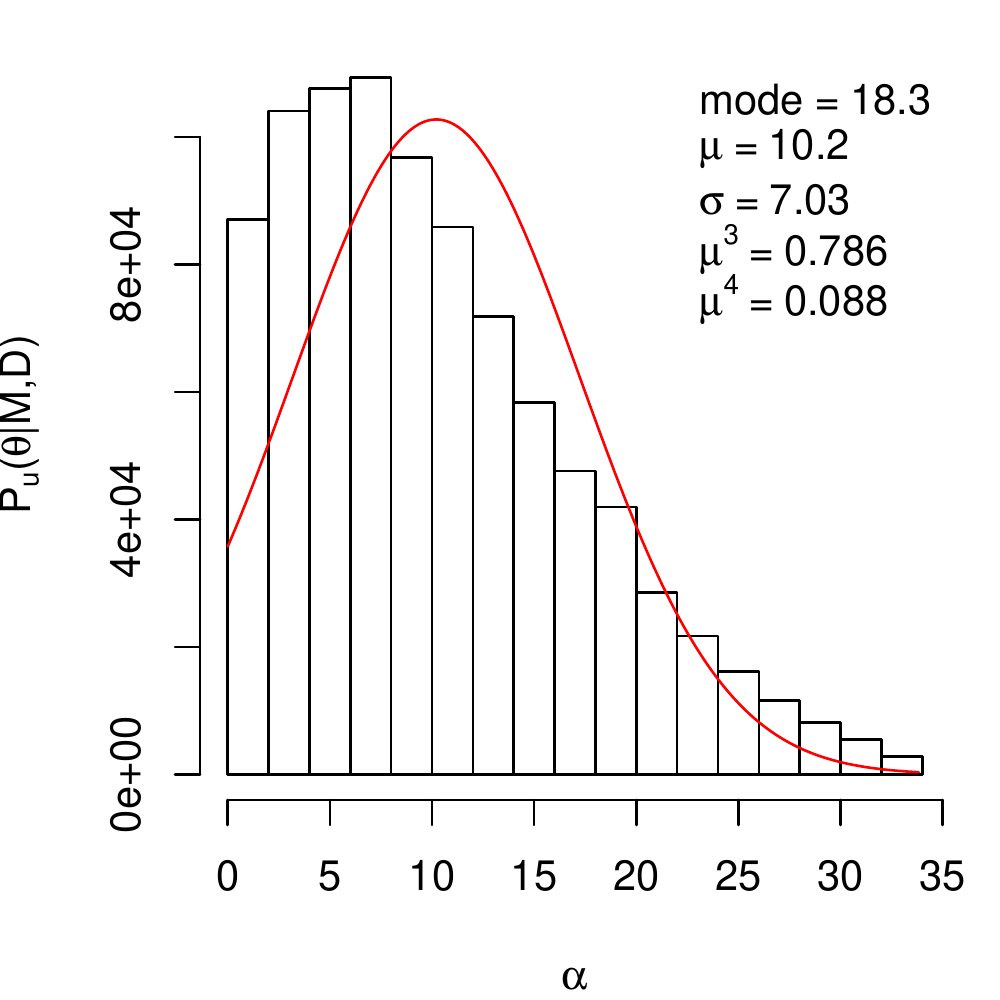}
  \includegraphics[scale=0.6]{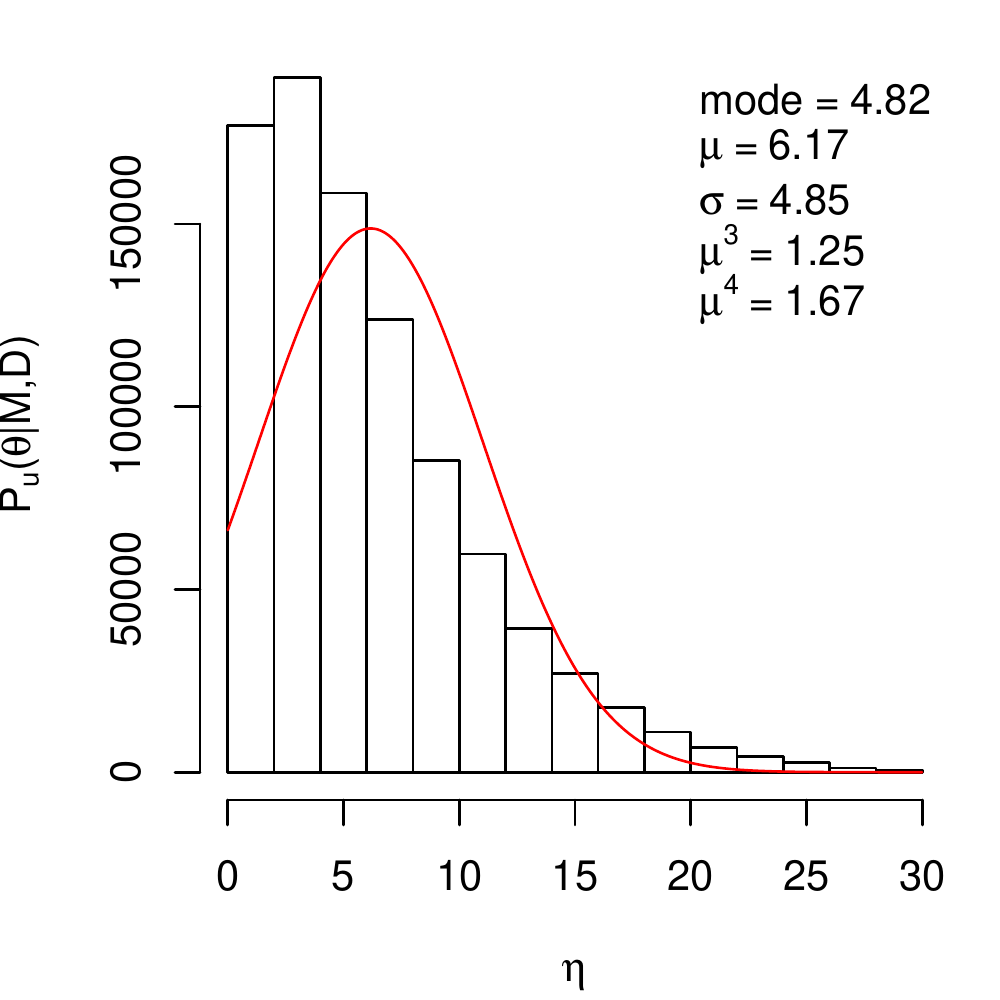}
\includegraphics[scale=0.6]{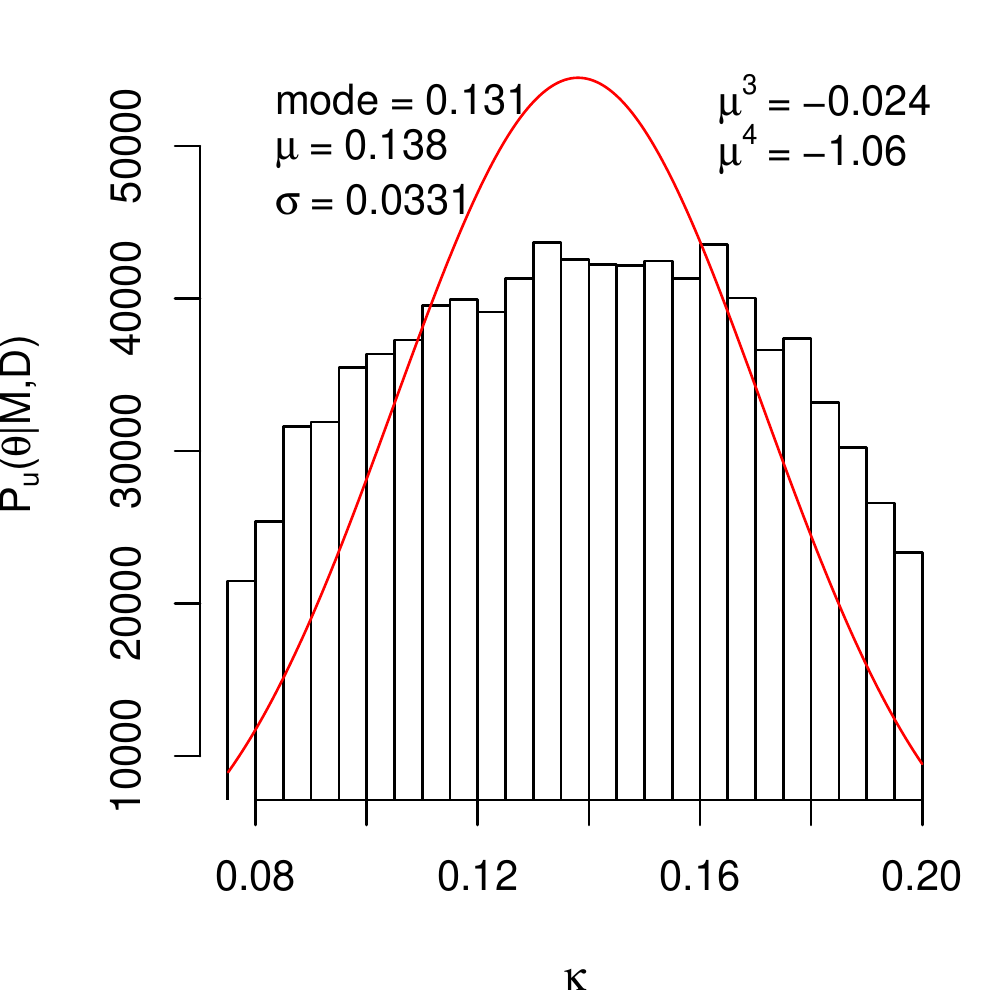}
\caption{The unnormalised posterior distribution of $s_w$, $\alpha$,
  $\eta$ and $\kappa$ of the LRJ1 model. The histograms are made with
  90,5004 posterior samples of a cold chain. For each panel, the red
  curve shows the fit of Gaussian distribution to the unnormalised
  posterior density. The mode, mean ($\mu$), standard deviation
  ($\sigma$), skewness ($\mu_3$) and kurtosis ($\mu_4$) are shown for
  each posterior distribution.}
\label{fig:post.extra}
\end{figure*}

Among the W1, RJ1 and LRJ1 models, the W1 model give similar results as the RJ1
and LRJ1 models, although RJ1 and LRJ1 can model a few data sets
slightly better due to adjusting extra free parameters. The LRJ1 model
is favoured by the GJ445 data sets with $(P,K)=$(20\,day, 6\,m/s) and
(40\,day, 4\,m/s). For the latter one, we show the posterior
distribution of parameters $\alpha$, $\eta$ and $\kappa$ of LRJ1 in
figure \ref{fig:post.extra}. We observe that the white noise $s_w$
dominates the total noise while the index dependent noises is
consistent with zero. For all the other data sets, we don't find
strong dependence of noise on the RHS index either. Despite this, the
RJ1 and LRJ1 models give fewer false positives than the white noise
model does, indicating a weak dependence of jitter on the RHS
index. Furthermore, the false positives detected by RJ1 and LRJ1
failed to pass the Bayes factor threshold. This is not caused by the
Bayesian penalization of model complexity because the Bayes factor of
RJ1 (or LRJ1) and RJ0 (or LRJ0) do not depend on the number of
parameters of the RJ (or LRJ) model. To test this further, we vary the Bayes factor threshold to see whether there is an optimal threshold which can reject more false positives and keep all true detections. But we failed to find such a value. For example, we increase the Bayes factor threshold to be 200, and apply this new threshold to test the signals detected by the white noise model. We find that the false positives for the GJ445 data set with $(P,K)=$ (40\,day, 4\,m/s) cannot be ruled out,
and the true signal in the GJ1 data set with $(P,K)=$ (20\,day, 2\,m/s)
is rejected. It means that we cannot confirm all true signals recovered by
the white noise model and reject false positives simultaneously by
adjusting the Bayes factor threshold. Considering these problems of
the white noise model and the complexity of the LRJ models, we
recommend the $R_{\rm HK}$-dependent jitter to model the excess noise in RV observations. 

Considering the limitations of different noise models, we set up
  a rule for modeling RV noise and selecting signals in order to avoid
  as many false positives and negatives as possible. We call this rule
``Goldilocks principle''.  Specifically, we suggest combining the
white noise model and the $R_{\rm HK}$-dependent jitter with the
moving average model in the following way to confirm
detections. First, we apply the three criteria introduced in section
\ref{sec:criteria} to confirm a signal detected by the model of $R_{\rm HK}$-dependent jitter. Then the signal is further confirmed if it
is also strong and unique (without local maxima exceeding the 10\% threshold, see
figure \ref{fig:post_artificial}) in the posterior distribution of the
moving average model. Finally, the signal is confirmed as a planet if it is not
found to be strong in the posterior distributions of the white noise model
(with zero eccentricity) for the activity indexes to avoid detecting activity-induced false positives that have a different phase in the RVs and activity indexes. 

\section{Discussions and conclusions} \label{sec:conclusions}
This work aims at comparing various noise models and inference
criteria for detecting weak signals in radial velocity data sets. We
define different noise models and introduce estimators of Bayes factor
to analyse artificial data sets. We find that the white noise model is
better than red noise models in detecting true signals. However,
the white noise model tends to interpret correlated noise as a signal,
and thus detect false positives. On the contrary, the red noise models, particularly the Gaussian process, usually interprets the signal as noise, at least partially, leading to false negatives. This is also the reason why the Gaussian process model is not favored even by the Gaussian process generated data sets (see table \ref{tab:controled}). This challenges the view that a simultaneous modeling of noise and signal components in data
would not result in overfitting or underfitting problems (e.g. \citealt{foreman-mackey15}). The solution of the problem is not only to perform modeling in the Bayesian framework but also to
properly model noise and signal according to a Goldilocks principle which could be obtained for each specific scientific question.

Comparing various Bayes factor estimators, we find that the BIC estimation of
Bayes factor combined with a Bayes factor threshold of 150 can reject
most false positives while other criteria either confirm more false
positives or reject a large proportion of true detections. In
addition, the truncated posterior mixture, harmonic mean and Deviance Information Criterion estimators do not converge properly. Meanwhile the Akaike Information Criterion and Chib's estimators penalize the one planet models too little and too much, leading to false positives and negatives, respectively. Given that all estimators of Bayes factors have short-comings \citep{ford07}, we adopt the BIC for practical reasons. 

We have applied the BIC-based signal detection threshold to analyze data sets with injected signals. We have simulated 27 data sets by injecting signals with various periods and amplitudes into the HARPS measurements of GJ1 and GJ361 and the Keck observations of GJ445. We find that the white noise model and the (lagged) $R_{\rm HK}$-dependent jitter models recover most injected signals. However, the Bayes factor threshold cannot reject all false positives found by the while noise model. Increasing the threshold cannot rule out all false positives and confirm all true detections simultaneously. On the contrary, the
Bayes factor threshold successfully reject all false positives detected
by the $R_{\rm HK}$-dependent noise models, although the dependence of jitter on
the $R_{\rm HK}$ is weak probably due to the low activity level of our
targets. To make the planet detection conservative, we suggest to form
a noise model framework by combining the $R_{\rm HK}$-dependent noise model
(RJ) and the moving average model. Since most planet hosts are M
dwarfs, our conclusions on modeling the RV noise of M dwarfs are
probably generic for exoplanet detections and so this work may also shed light on the noise modeling for hotter stars. 

We also test the sensitivity of the evidences for red noise models to
their kernels, and don't find any significant improvement for the test
cases analysed here. Since
the injection data sets are from different instruments and with different sizes,
our quantification of the limitations of various noise models are
probably generic for detecting planets in RV observations. Our results
indicate that flexible noise models such as Gaussian
processes may underestimate the number of Keplerian signals. This is
supported by \citet{tuomi12}'s choice of first order moving average model to
reduce jitter rather than higher order moving average models. In addition, the
Tuomi et al. group ``won'' the RV Challenge using the moving average model while
other groups failed to recover as many signals using more flexible models. 
This is consistent with our findings that the usage of flexible noise
models tend to result in false negatives when the models do not correctly reflect the underlying physics of stellar activity. 

This difference between noise models is also evident from the
controversy over the validation of the number of planets, where red
noise models find less signals than other models, e.g. GJ 581 and GJ
667C discussed in the introduction section. These controversies are consistent with
our conclusion that red noise models lead to false negatives while the
white (or $R_{\rm HK}$-dependent) noise model lead to false positives. To
avoid both, we define a Goldilocks principle by combining the $R_{\rm HK}$-dependent noise model with the moving average model and a BIC-based
signal detection criterion. This principle also provides a clue for
noise modeling in other fields. For example, stochastic models may not
be appropriate for modeling the glacial-interglacial cycles over the
Pleistocene because they tend to give false negatives. This can be investigated
through injecting Earth's orbital variations into noisy climate data and recovering
them using stochastic noise models in combination with orbital
models. Another example is the detection of periodic signals in quasar light curves. The optical variability of quasars could be caused by random processes, rotations of binary black holes, uneven sampling and/or correlated noise. Since RV variations have similar characteristics, our work may also provide insights for disentangling periodic signals from stochastic variability in quasar light curves (e.g. \citealt{graham15,vaughan16}). 

In summary, the Goldilocks principle provides an approach to balance between
overfitting and underfitting of noise by statistical models, which may
poorly reflect the underlying physics and thus are unable to
disentangle noise from signals. Although a Gaussian process framework has been proposed to partly account for the underlying physics \citep{rajpaul15}, the jitter
may not be properly modeled due to the flexibility of Gaussian process models and
the simplification of the complex relationship between RV variations
and stellar activity indexes. Further studies on the statistical property of
stellar activity proxies and their connection with RV variations
are essential steps towards an astrophysically motivated modeling of
stellar jitter. A probable method is the nonlinear time series analysis which
connects the nonlinear dynamical system with the time series of
some system outputs \citep{kantz04,sugihara12}. This idea has inspired
us to build the lagged $R_{\rm HK}$-dependent jitter model which performs well in our analyses. Moreover, correlated noise and deterministic signals can be well distinguished using surrogate time series, a concept developed in the community of nonlinear time series
\citep{schreiber00}. These facts justify further investigations into the nonlinear approach of modeling RV noise. 

\section*{Acknowledgements}
FF, MT and HJ are supported by the Leverhulme Trust (RPG-2014-281) and
the Science and Technology Facilities Council (ST/M001008/1). We
used the ESO Science Archive Facility to collect radial velocity data sets. We also thank the referee, David Kipping, for valuable comments. 

\section*{Appendix: Keck data of GJ 445}
\begin{table*}
\centering 
\caption{The Keck measurements of GJ 445. }
\label{tab:prior}
  \begin{tabular}  {c*{3}{c}}
Julian Days&RV[m/s]&RV error[m/s]&$R_{\rm HK}$\\\hline
2450840.15414&1.82& 2.52&0.7041\\
2450862.05138&-15.10& 2.61&0.4365\\
2451171.11243&6.22& 3.09&0.5667\\
2451173.15170& -6.46& 2.70&0.5500\\
2451174.14020&3.09& 2.91&0.7017\\
2451229.02528&-16.97& 2.88&0.6536\\
2451312.83299&4.72& 2.33&0.6772\\
2451581.05520& -0.05& 2.72&0.6771\\
2451702.84804&5.85& 2.61&0.5760\\
2451983.03906& -3.28& 3.36&0.3824\\
2452333.05276& 10.70& 3.55&0.5765\\
2452654.07529& -2.00& 2.94&0.3798\\
2452681.03126& 11.68& 3.64&0.4286\\
2453018.11966&7.61& 2.76&0.3705\\
2453399.01632&0.80& 2.87&0.6102\\
2454131.08186&7.17& 3.03&0.6257\\
2454277.80190& -0.04& 2.59&0.5708\\
2454278.82143& -3.37& 2.88&0.5773\\
2454279.81260& -1.15& 2.79&0.6036\\
2454285.83863&4.57& 3.37&0.5528\\
2454294.86588&6.86& 3.01&0.7721\\
2454304.84650&5.53& 2.55&0.4551\\
2454305.85101&0.00& 2.30&0.5399\\
2454306.84962& -4.73& 2.59&0.6472\\
2454307.88909& -4.51& 2.74&0.2917\\
2454308.86895& -6.74& 3.57&0.3674\\
2454309.85157& -0.65& 2.87&1.2438\\
2454310.84423& -2.85& 2.76&0.5632\\
2454311.83910&-13.70& 2.95&0.5975\\
2454312.83456& -4.73& 2.87&0.9761\\
2454313.83837&1.69& 2.89&0.5117\\
2454314.86490& -0.83& 3.16&0.6875\\
2454455.14057&5.31& 2.97&0.7055\\
2454455.14674&2.30& 2.99&0.7308\\
2454491.04532&0.12& 2.54&0.5859\\
2454546.04764&0.62& 2.33&0.5356\\
2454600.95556& -0.47& 2.74&0.6064\\
2454601.92483& -2.04& 1.99&0.6463\\
2454701.74852&2.92& 2.85&0.4183\\
2454701.75598&4.35& 2.91&0.5082\\
2454702.74186&0.80& 2.87&0.7380\\
2454702.74938& -1.07& 3.21&0.5221\\
2454703.74693&4.92& 2.61&0.3905\\
2454703.75477& -0.73& 2.59&0.6241\\
2454704.73349&7.15& 2.83&0.6441\\
2454704.74171& -0.16& 3.44&0.4719\\
2454968.95856& -2.12& 2.71&0.6697\\
2454968.96572& -0.99& 3.23&0.4139\\
2455021.85243& -3.12& 3.86&0.4572\\
2455049.79414& -4.34& 3.01&0.6463\\
2455258.99615&-12.23& 2.82&0.4616\\
2455313.90921&9.88& 2.59&0.5754\\
2455371.79182&0.60& 2.68&0.3061\\
2455637.04757& -3.31& 2.43&0.5640\\
2455638.02685&1.23& 2.56&0.4916\\
2455663.86110&0.46& 2.30&0.6717\\
2455668.88064& 10.32& 2.40&0.6443\\
2455670.92034&6.46& 2.45&0.6590\\
2455671.92802&5.80& 2.60&0.7044\\
2455672.92840&4.46& 2.45&0.7519\\
2455673.89382&6.65& 2.63&0.6778\\
2455703.84580& -6.13& 2.80&0.5339\\
2455704.76119& -5.27& 2.78&0.5667\\
2455705.77066& -7.64& 2.52&0.6797\\\hline
  \end{tabular}
\end{table*}
\bibliographystyle{mn2e}
\bibliography{nm}  
\end{document}